%% file: lineno-lipics-v2021-sample-article.tex
\PassOptionsToPackage{UKenglish}{babel}
\PassOptionsToClass{cleveref,autoref,thm-restate}{lipics-v2021}
\documentclass[a4paper]{lipics-v2021}

\pdfoutput=1 

\usepackage{lmodern} 
\usepackage[T1]{fontenc}
\usepackage{color}
\usepackage{amsmath, amssymb, amsthm}
\usepackage{mathtools}
\usepackage{bbm}
\usepackage{dsfont}
\usepackage{graphics}
\usepackage{diagbox}
\usepackage{pifont}
\usepackage{tikz}
\usepackage{bbm, bm}

\usetikzlibrary{arrows.meta}
\usepackage{environ}
\usepackage{framed}
\usepackage{url}
\usepackage[linesnumbered,ruled,vlined]{algorithm2e}
\usepackage[noend]{algpseudocode}
\usepackage[labelfont=bf]{caption}
\usepackage{framed}
 \usepackage[framemethod=tikz]{mdframed}
\usepackage{appendix}
\usepackage{graphicx}
\usepackage[textsize=tiny]{todonotes}
\usepackage{tcolorbox}
\allowdisplaybreaks[1]
\usepackage{nicefrac}
\usepackage{thm-restate}

\usepackage{array}
\usepackage{multirow}

\newtheorem*{rem*}{Remark}

\theoremstyle{definition}
\newtheorem{problem}{Problem}

\newcommand{\occ}{\mathtt{occ}}
\newcommand{\cnt}{\mathtt{cnt}}
\newcommand{\pos}{\mathtt{pos}}

\newcommand{\BSS}{\mathtt{BASS}}

\newcommand{\ext}{\mathtt{ext}}

\newcommand{\rev}{\mathsf{rev}}
\usetikzlibrary{arrows}
\tikzset{edge/.style={->,> = latex'}}
\tikzset{special/.style={fill=red!50,circle,minimum size=0.6cm,inner sep=1pt}}

\title{Nearly Optimal Internal Dictionary Matching}

\author{Jingbang Chen\footnote{Part of this work was done while at Georgia Tech and the University of Waterloo.}}{The Chinese University of Hong Kong, Shenzhen, China \and \url{https://chenjb1997.github.io/}}{chenjb@cuhk.edu.cn}{https://orcid.org/0000-0002-7279-0801}{}

\author{Jiangqi Dai}{Massachusetts Institute of Technology, USA}{jqdai@mit.edu}{}{}

\author{Qiuyang Mang}{The Chinese University of Hong Kong, Shenzhen, China \and \url{https://joyemang33.github.io/}}{qiuyangmang@link.cuhk.edu.cn}{}{}

\author{Qingyu Shi}{Feicheng Hailiang Foreign Language School, China}{qingyuqwq@gmail.com}{}{}

\author{Tingqiang Xu}{Tsinghua University, China}{xtq23@mails.tsinghua.edu.cn
}{}{}

\authorrunning{J. Chen, J. Dai, Q. Mang, Q. Shi, and T. Xu}

\Copyright{Jingbang Chen, Jiangqi Dai, Qiuyang Mang, Qingyu Shi, and Tingqiang Xu}



\category{} 

\relatedversion{}

\nolinenumbers

\EventEditors{Philipp Bille, Seth Pettie, and Sabine Storandt}
\EventNoEds{3}
\EventLongTitle{34th Annual European Symposium on Algorithms (ESA 2026)}
\EventShortTitle{ESA 2026}
\EventAcronym{ESA}
\EventYear{2026}
\EventDate{August 31--September 4, 2026}
\EventLocation{L'Aquila, Italy}
\EventLogo{}
\SeriesVolume{388}
\ArticleNo{60}

\begin{document}

\maketitle

\begin{abstract}
We study the \textit{internal dictionary matching} (IDM) problem where a dictionary $\mathcal{D}$ containing $d$ substrings of a text $T$ over a linearly sortable alphabet is given, and each query concerns the occurrences of patterns in $\mathcal{D}$ in another substring of $T$.
We propose a novel $O(n)$-sized data structure named \textit{Basic Substring Structure} (BASS) where $n$ is the length of the text $T.$ With BASS, we are able to handle \textit{all types of queries} in the IDM problem in nearly optimal query and preprocessing time. Specifically, our results include:
\begin{itemize}
    \item The first algorithm that answers the \textsc{CountDistinct} query in $\Tilde{O}(1)$ time with $\Tilde{O}(n+d)$ preprocessing, where we need to compute the number of distinct patterns that exist in $T[l,r]$. Previously, the best result was $\Tilde{O}(m)$ time per query after $\Tilde{O}(n^2/m+d)$ or $\Tilde{O}(nd/m+d)$ preprocessing, where $m$ is a chosen parameter.
    \item Faster algorithms for two other types of internal queries. We improve the runtime for \textbf{(1)} \textit{Occurrence counting} (\textsc{Count}) queries to $O(\log n/\log\log n)$ time per query with $O(n+d\sqrt{\log n})$ preprocessing from $O(\log^2 n/\log\log n)$ time per query with $O(n\log n/\log \log n+d\log^{3/2} n)$ preprocessing. \textbf{(2)} \textit{Distinct pattern reporting} (\textsc{ReportDistinct}) queries to $O(1+|\text{output}|)$ time per query from $O(\log n+|\text{output}|)$ per query.
\end{itemize}

In addition, we match the optimal runtime in the remaining two types of queries, \textit{pattern existence} (\textsc{Exists}), and \textit{occurrence reporting} (\textsc{Report}). We also show that BASS is more generally applicable to other internal query problems.
\end{abstract}

\input{intro}

\input{main}

\bibliography{refs}

\appendix

\input{more}

\input{proof}

\end{document}

%% file: intro.tex
\section{Introduction}

The \textit{pattern matching}, as well as the \textit{dictionary matching} problems, have been extensively studied for decades, where we need to preprocess a dictionary $\mathcal{D}$ containing one or multiple patterns to answer their occurrences in each query text. In addition to classical methods~\cite{knuth1977fast,aho1975efficient}, there are also studies on developing approximate algorithms \cite{baeza1998fast,okazaki2010simple,boytsov2011indexing,Cislak_Grabowski_2017}, and algorithms in the dynamic dictionary setting \cite{amir1994dynamic,amir1995improved,sahinalp1996efficient,chan2005dynamic,hon2009succinct}. 

In this paper, we consider the \textit{internal} setting where each pattern or query is represented by a substring of a given large text $T$. 
Such an ``internal'' concept is not new in string algorithms. For example, the famous suffix array algorithm~\cite{manber1993suffix,gonnet1992new,karkkainen2003simple,nong2009linear} sorts all suffixes of $T$, which helps to answer multiple types of queries such as the longest common prefix (LCP) of two suffixes. In this case, all related texts (suffixes) are internal with respect to $T$. Computing the LCP of any two suffixes can be easily generalized to computing the LCP of any two substrings (fragments), known as the \textit{longest common extension} (LCE) queries introduced by Landau and Vishkin~\cite{landau1988fast}, which could be one of the earliest internal queries. 

In recent years, the \textit{internal pattern matching} (IPM) problem has been widely studied, which requires us to compute the occurrences of a substring of $T$ in another substring of $T$. Recent research has developed algorithms that run in sublogarithmic time and occupy near-linear space \cite{keller2014generalized,kociumaka2014internal,kociumaka2019efficient}. Benefit from the IPM algorithms, many internal queries have been studied, such as shortest unique substrings~\cite{abedin2020efficient}, shortest absent string~\cite{badkobeh2022internal}, BWT substring compression~\cite{babenko2014wavelet}, circular pattern matching~\cite{iliopoulos2023linear}, etc. Moreover, the IPM queries have been used in some other problems, including approximate pattern matching~\cite{charalampopoulos2020faster,charalampopoulos2022faster}, approximate circular pattern matching~\cite{charalampopoulos2021circular,charalampopoulos2022approximate}, RNA folding~\cite{das2021improved}, and computing string covers~\cite{radoszewski2020efficient}.

Similar to dictionary matching, a natural generalization of the IPM problem is the newly proposed \textit{internal dictionary matching} (IDM) problem, where we consider multiple patterns. It was first introduced and studied by Charalampopoulos et al.~\cite{charalampopoulos2020counting,charalampopoulos2021internal}. In such a setting, given a large text $T$ of length $n$ in advance, we consider a dictionary $\mathcal{D}$ consisting of substrings of $T$. We denote the number of patterns as $d$.
The following types of internal queries are studied in the IDM problem:
\begin{itemize}
    \item $\textsc{Exists}(l,r)$: Decide whether at least one pattern from $\mathcal{D}$ occurs in $T[l,r]$.
    \item $\textsc{Report}(l,r)$: Report all occurrences of all patterns from $\mathcal{D}$ in $T[l,r]$.
    \item $\textsc{ReportDistinct}(l,r)$: Report all distinct patterns from $\mathcal{D}$ in $T[l,r]$.
    \item $\textsc{Count}(l,r)$: Count the number of all occurrences of all patterns from $\mathcal{D}$ in $T[l,r]$.
    \item $\textsc{CountDistinct}(l,r)$: Count the number of distinct patterns from $\mathcal{D}$ in $T[l,r]$.
\end{itemize}
Compared to the classical dictionary matching problem, the IDM problem is especially applicable to any internal dictionary whose total length of all patterns (substrings) $L$ is much larger than the number of patterns $d$, such as the dictionary of all palindrome substrings, all square substrings, and all non-primitive substrings. Combined with algorithms that efficiently extract the endpoints of patterns~\cite{groult2010counting,crochemore2014extracting,bannai2016computing}, an IDM algorithm can answer queries regarding such an internal dictionary in time and space \textit{not} depending on $L.$
IDM algorithms for the palindromes case~\cite{rubinchik2017counting} and the squares case~\cite{charalampopoulos2021internal} have already been developed. 
Same as IPM, IDM also has many practical applications. For example, in bioinformatics, the multi-pattern matching query has many deployed applications, most of which are under the internal setting and can be reduced to IDM~\cite{tevatia2015multi,zhou2016efficient}.

Previous work~\cite{charalampopoulos2020counting} shows that the product of the time to process an update to $\mathcal{D}$ and the time to answer an internal query cannot be $O(n^{1-\epsilon})$ for any constant $\epsilon>0$, unless the Online Boolean Matrix-Vector Multiplication conjecture~\cite{henzinger2015unifying} is false. This conditional lower bound demonstrates the hardness of efficient algorithms in the dynamic dictionary case. In this paper, we focus mainly on the case of a static dictionary.

\subsection{Our Results} 

\renewcommand{\arraystretch}{1.5}

\begin{table}
\centering
\caption{Algorithmic Results}
\label{table:result}

\resizebox{0.98\linewidth}{!}{
\begin{tabular}{|c|c|c|c|c|c|} 
    \hline
    \multirow{2}{*}{\diagbox[width=9em, height=3.6em]{}{}}           & \multicolumn{3}{c|}{\textbf{Our results}}                                                                    & \multicolumn{2}{c|}{\textbf{Previous results}}                                                                           \\ 
    \cline{2-6}
                                            & \textbf{Preprocessing}                                  & \textbf{Query time}                     & \textbf{Section}           & \textbf{Preprocessing}                                            & \textbf{Query time}                                           \\ 
    \hline
    \textsc{Exists}                         & $O(n+d)$                                       & $O(1)$                         &    \ref{sec:e-r}                & $O(n+d)$                                                 & $O(1)$                                               \\ 
    \hline
    \textsc{Report}                         & $O(n+d)$                                       & \textsuperscript{1}$O(1+x)$             &  \ref{sec:e-r}                 & $O(n+d)$                                                 & \textsuperscript{1}$O(1+x)$                                             \\ 
    \hline
    \textsc{Count}                          & $O(n+d\sqrt{\log n})$                          & $O(\frac{\log n}{\log\log n})$ &   \ref{sec:count}                & $O(\frac{n\log n}{\log \log n}+d\log^{3/2} n)$           & $O(\frac{\log^2 n}{\log\log n})$                     \\ 
    \hline
    \textsc{ReportDistinct}                 & $O(n\log n +d)$                                      & \textsuperscript{1}$O(1+x)$              &  \ref{sec:reportdistinct}                 & $O(n\log n+d)$                                           & \textsuperscript{1}$O(\log n+x)$                     \\ 
    \hline
    \multirow{3}{*}{\textsc{CountDistinct}} & \multirow{3}{*}{$O(n\log^2 n+d\sqrt{\log n})$} & \multirow{3}{*}{$O(\log n)$}   & \multirow{3}{*}{\ref{sec:countdistinct}} & \textsuperscript{2}$\tilde{O}(n^2/m + d)$                                   & \textsuperscript{2}$\tilde{O}(m)$                                       \\ 
    \cline{5-6}
                                            &                                                &                                &                   & \textsuperscript{2}$\tilde{O}(nd/m+d)$                                      & \textsuperscript{2}$\tilde{O}(m)$                                       \\ 
    \cline{5-6}
                                            &                                                &                                &                   & \textsuperscript{3}$O(n\log^{1+\epsilon}n+d\log^{3/2}n)$ & \textsuperscript{3}$O(\frac{\log^2 n}{\log\log n})$  \\
    \hline
\end{tabular}
}
\vspace{1em}
\renewcommand\TPTtagStyle[1]{\textsuperscript{#1}}
\begin{tablenotes}
   \footnotesize
   \item \textsuperscript{1} $x$ denotes the length of output.
   \item \textsuperscript{2}  $m$ can be arbitrary positive integer
   \item \textsuperscript{3} 2-approximate
 \end{tablenotes}
\end{table}

In this paper,  we propose a new $O(n)$-sized data structure named \textit{Basic Substring Structure (BASS)} that organizes $T$'s substrings in a novel way (\Cref{sec:bss}). BASS builds upon the idea of equivalence classes of substrings, an approach that has been explored in previous work, such as the seminal work by Blumer et al.~\cite{blumer1987complete,crochemore1997direct}, which introduced the concept of equivalence classes in the context of the compact acyclic word graph (CDAWG).
While the equivalence class concept is indeed not new, the key contribution of this paper lies in the grid structure that organizes these equivalence classes in a two-dimensional grid. This novel structure enables the efficient handling of substring queries, something that earlier works did not fully address. In contrast to traditional representations like the CDAWG, which represent equivalence classes in a graph-like structure, the grid-based approach in BASS allows for more efficient indexing and querying of patterns.

BASS is the first data structure that is able to handle \textit{all} types of queries of the IDM problem in nearly optimal query and preprocessing time. This contrasts with previous works where they treated different queries with tailored approaches. 
\Cref{table:result} gives an overview of our BASS-based 
 algorithms on different query types compared to previous results. All algorithms take $\Tilde{O}(n+d)$\footnote{The $\Tilde{O}(\cdot)$ notation suppresses $\log^{O(1)} n$ factors for inputs of size $n$.} space. 
 
\subsubsection{Counting Distinct Patterns} By introducing an extra parameter $m$, the previous best result answers \textsc{CountDistinct} queries in $\Tilde{O}(m)$ time after $\Tilde{O}(n^2/m+d)$ or $\Tilde{O}(nd/m+d)$ preprocessing~\cite{charalampopoulos2021internal}. The question of whether any data structure answers \textsc{CountDistinct} queries exactly in $\Tilde{O}(1)$ time each after $\Tilde{O}(n+d)$ preprocessing was open. As one of the main contributions, we answer this question affirmatively:

\begin{restatable}[CountDistinct]{theorem}{countdistinct}
\label{CD-res}
\textsc{CountDistinct}$(i, j)$ can be answered in $O(\log n)$ time with a data structure of $O(n\log^2 n + d)$ size that can be constructed in $O(n\log^2 n + d\sqrt{\log n})$ time.
\end{restatable}

Similar to~\cite{charalampopoulos2020counting}, for the case of a dynamic dictionary, where patterns can be added or deleted, we can construct algorithms that process updates in $\Tilde{O}(n^\alpha)$ time and answer queries \textsc{CountDistinct}$(l,r)$ in $\Tilde{O}(n^{1-\alpha})$ time for any $0<\alpha<1$. For this type of query, this is the first result that matches the known conditional lower bound up to subpolynomial factors.

\subsubsection{Other Internal Queries} As in previous work~\cite{charalampopoulos2020counting}, \textsc{Exists}, \textsc{Report}, \textsc{ReportDistinct}, \textsc{Count} can be answered in $\Tilde{O}(1+|\text{output}|)$ time after $\Tilde{O}(n+d)$ preprocessing of $T$. Our BASS-based algorithms match all previous results and achieve some improvements.

For the \textsc{Count} query, previous works take $O(\frac{\log^2 n}{\log\log n})$ per query after $O(\frac{n\log n}{\log \log n}+d\log^{3/2} n)$ preprocessing. Our result improves the run time for both query and preprocessing:

\begin{restatable}[Count]{theorem}{counttheorem}
\label{count2}
    \textsc{Count}$(l, r)$ can be answered in $O(\frac{\log n}{\log \log n})$ time with a data structure of size $O(n + d)$ that can be constructed in $O(n+d\sqrt{\log n})$ time.
\end{restatable}

Also, we prove that there is an unconditional lower bound on \textsc{Count}$(l,r)$, stating that $O(\frac{\log n}{\log \log n})$ query time is optimal for all data structures of size $\tilde{O}(n)$ \cite{Pat07}. Thus, our query time for $\textsc{Count}(l,r)$ queries is optimal.

For \textsc{ReportDistinct} queries that output all distinct patterns that occurred, using the same preprocessing time, our query time improves to $O(1+|\text{output}|)$ from $O(\log n +|\text{output}|)$:

\begin{restatable}[ReportDistinct]{theorem}{reportdistinct}
\textsc{ReportDistinct}$(l,r)$ can be answered in $O(1+x)$ time with a data structure of size $O(n + d)$ that can be constructed in $O(n\log n + d)$ time, where $x$ denotes the length of output.    
\end{restatable}

For \textsc{Exists}, \textsc{Report} queries, although taking different approaches, both our algorithm and previous works need $O(n+d)$ preprocessing time and $O(1+|\text{output}|)$ query time (\cref{theo:exist}, \cref{theo:report}). 
In addition, BASS can find further applications, particularly for various kinds of internal queries. We discuss the \textit{range longest common substring (RLCS)} problem as a showcase. Specifically, given two strings $S$ and $T$, each query \textsc{LCS}$(l,r)$ requires computing the longest common substring between $S$ and $T[l,r]$. With BASS, we can construct a data structure with $O(n)$ preprocessing time and constant query time, achieving optimality (\cref{theo:rlcs}) and improving the previous best result of $O(\log n)$ query time~\cite{amir2020dynamic}.


\section{Preliminaries}
\label{sec:prelim}
We first give some common definitions and notations for strings, aligned with \cite{crochemore2007algorithms}. Let $T=T_1T_2T_3\cdots T_n$ be a string of length $|T|=n$ over a linearly sortable alphabet $\Sigma$. We refer to elements in $\Sigma$ as letters or characters. For any two indexes $1 \leq l \leq r \leq n$, the string constructed by concatenating $T_l,T_{l+1},\cdots,T_r$ is denoted as $T[l,r]$.

\paragraph{Substrings, prefixes, and suffixes.}
For two strings $T$ and $t$, $t$ is a substring of $T$ if and only if there exist some indexes $1 \leq l \leq r \leq |T|$ such that $T[l,r]=t$. We can also say that $t$ occurs in $T$ or $T$ contains $t$. A prefix of $T$ is a substring of the form $T[1, i]$ for some $1 \leq i \leq |T|$. Similarly, a suffix of $T$ is a substring of the form $T[i, |T|]$.

\paragraph{Occurrence sets.}
Given a string $T$, for any pattern $t$, $\occ_{T}(t)$ is defined as the set of distinct index pairs $(l,r)$ $(1 \leq l \leq r \leq n)$ such that $T[l,r]=t$, indicating the occurrences of $t$ in $T$. If the dictionary string is not ambiguous, we can also write $\occ(t)$.

A \textbf{trie} is a tree data structure where each edge is labeled by a single character, with paths from the root to specific nodes forming complete words. For a set $S$ of strings, we denote the trie built by $S$ as $\mathcal{E}(S)$. Also, we denote $\texttt{str}'(u)$ as the string formed by the path from the root to $u$. If $\texttt{str}'(u)\in S$, we call $u$ a terminal node.


For a string $T$ over a linearly sortable alphabet $\Sigma$, the \textbf{suffix tree} $\mathcal{T}(T)$ is a compact trie built by all suffixes of $T$. If a node in trie has only one child and is not a terminal node, the node is compressed. Thus, edges are labeled with character sequences instead of a single character. We denote $\texttt{str(u)}$ as the set of substrings $\texttt{str}'(v)$ for all $v$ that is compressed into $u$. Thus, the union of all $\texttt{str}(u)$ is all different substrings of $T$, and we say a node $u$ recognizes a substring $t$ if $t\in \texttt{str}(u)$. Also, it is proved that substrings in $\texttt{str}(u)$ have the same set of start positions. Additionally, for non-root node $u$, we denote $\texttt{parent}(u)$ to be the parent of $u$, and by definition, any string in $\texttt{str}(\texttt{parent}(u))$ is a prefix of all strings in $\texttt{str}(u)$. Lastly, we denote $\texttt{maxstr}(u)$ as longest string stored in $\texttt{str}(u)$ and $\texttt{len}(u)$ as the length of $\texttt{maxstr}(u)$.

\begin{theorem}[\cite{farach2000sorting}]
For a given string $T$ over a linearly sortable alphabet $\Sigma$, the size of suffix tree $\mathcal{T}(T)$ is $O(|T|)$, and it can be constructed in $O(|T|)$ time.
\end{theorem}

\begin{figure}[ht]
    \centering
    \includegraphics[width=0.65\linewidth]{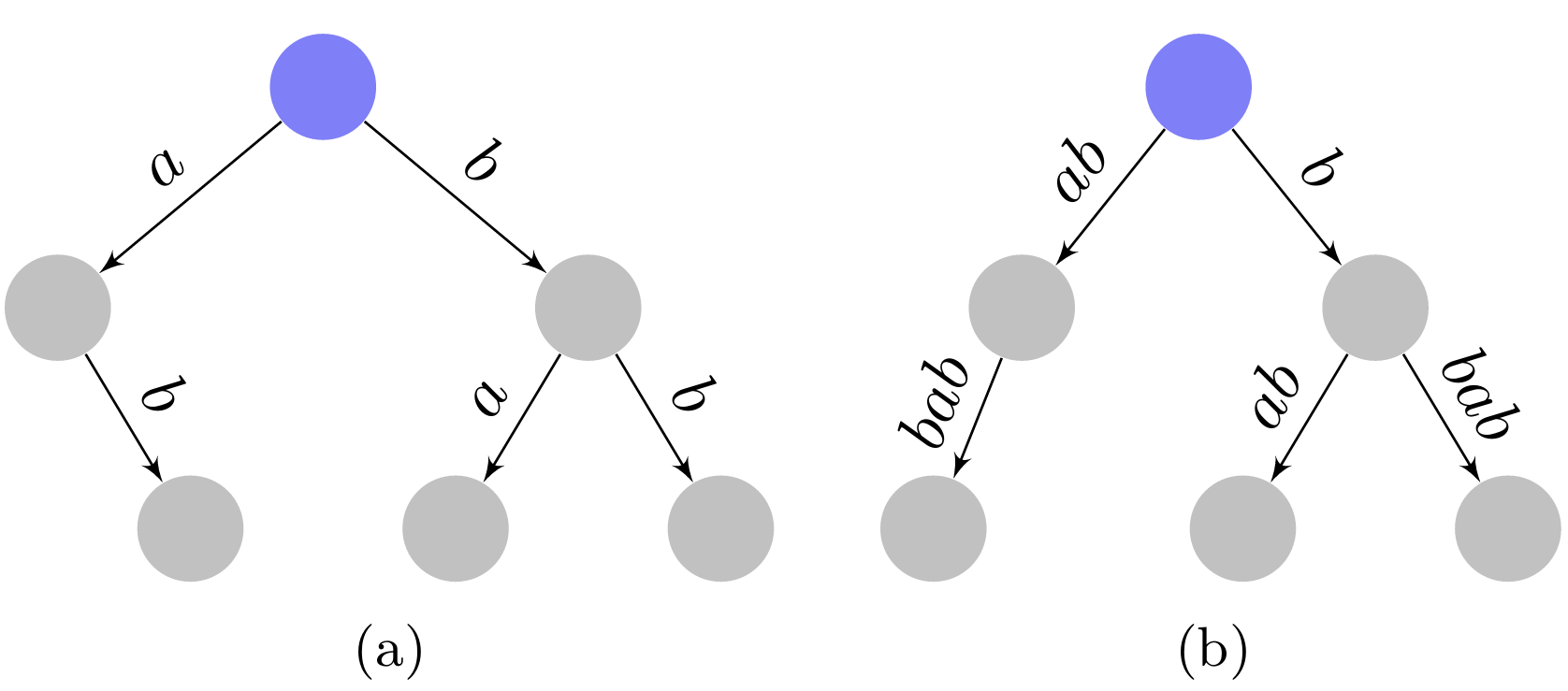}
    \caption{Two illustrative examples: \textbf{(a)} $\mathcal{E}(\{\texttt{ab}, \texttt{ba}, \texttt{bb}\})$; \textbf{(b)} $\mathcal{T}(\texttt{abbab})$.}
\end{figure}

%% file: main.tex
\section{Basic Substring Structure}
\label{sec:bss}
\subsection{Grid}
\label{sec:gridsystem}
\begin{figure}[ht]
    \centering
    \includegraphics[width=1\linewidth]{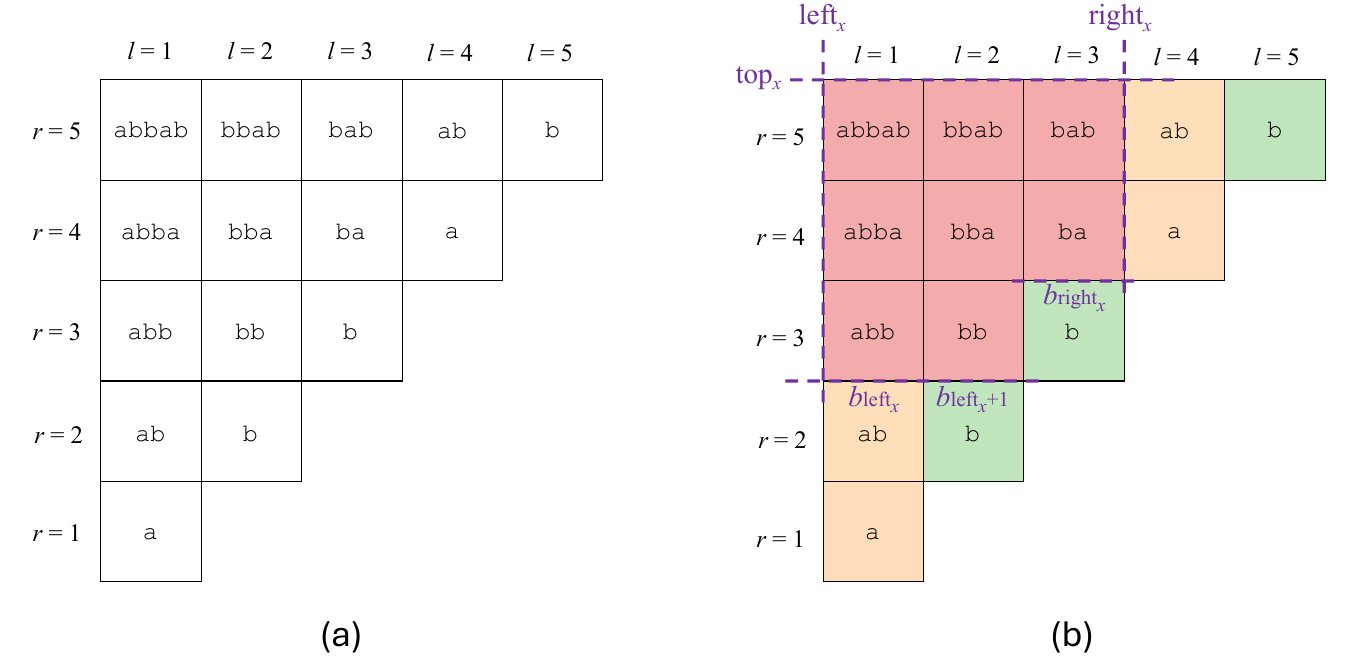}
    \caption{\textbf{(a)} The grid of the string $\texttt{abbab}$. \textbf{(b)} The blocks correspond to each equivalence class of the string, where each color corresponds to the blocks of an equivalence class. }
    \label{figure:bss-blocks}
\end{figure}

The \textit{Basic Substring Structure} (BASS) of a string $T$, denoted by $\BSS(T)$, is based on a two-dimensional grid $G$ which maps every substring $T[l, r]$ to a point $(l, r)$. Each row $i$ corresponds to all the substrings of $T$ ending at position $i$, and each column $j$ corresponds to all the substrings of $T$ starting at position $j$. \Cref{figure:bss-blocks} (a) gives an example with $T=\texttt{abbab}$.

Recall that $\textsc{Count}(l,r)$ computes the number of all occurrences of all patterns in $T[l,r]$. For every substring $T[l,r]$, we mark its corresponding point $(l,r)$ if and only if $T[l,r] \in \mathcal{D}$. The answer to $\textsc{Count}(l,r)$ is equivalent to the number of marked points within the rectangle $[l, \infty] \times [-\infty, r]$. However, since there could be up to $O(nd)$ marked points, it is inefficient to preprocess them one by one. 

\subsection{Equivalence Class}

Now, we show how these points can be efficiently stored. To do this, we need to introduce the concept of equivalence classes in the BASS. We begin by defining an extension operation for any substring $t$:

\begin{definition}[Extension]
    For any substring $t$ of string $T$, $\ext(t)$ is defined as the longest string $s$ such that \textbf{(i)} $t$ is a substring of $s$, and \textbf{(ii)} $|\occ(t)|=|\occ(s)|$.
\end{definition}

The extension operation is well defined because, for any substring $t$, there is only one longest such string $s$. The proof of this uniqueness can be found in \cref{sec:proof}. 

For any two substrings $t_1$ and $t_2$ of $T$, we define them as equivalent if and only if $\ext(t_1)=\ext(t_2)$. That is to say, substrings $s$ that share the same value of $\ext(s)$ are classified into the same equivalence class.

\begin{remark}
    The concept of extension is commonly referred to in the literature as maximal repeats, and the equivalence class defined here corresponds exactly to the nodes of the compact acyclic word graph (CDAWG) \cite{blumer1987complete,crochemore1997direct}. However, in the subsequent discussion, we explore the substrings within a single equivalence class in more detail, uncovering a more refined structure that extends the CDAWG.
\end{remark}

Now, we will map the equivalence classes to the grid $G$. To illustrate, we assign a unique color to each equivalence class $C$ and color each cell $(i, j)$ as the color of the equivalence class of $T[i, j]$. Then, we call each maximal connected component with a single color in the grid as a \textit{block}, which exhibits a staircase-like shape as shown in \Cref{figure:bss-blocks} (b).
More formally, we have the following lemma.

\begin{lemma}
\label{lemma:blockshape}
    For any block $B$, it can be described by using the following parameters:
    \begin{itemize}
        \item Three integers $\texttt{left}_B, \texttt{right}_B, \texttt{top}_B$;
        \item A non-decreasing integer sequence $b_{\texttt{left}_B} \leq b_{\texttt{left}_B+1} \leq \ldots \leq b_{\texttt{right}_B}$;
    \end{itemize}
    Specifically, we have $B = \{(i, j)\mid \texttt{left}_B \leq i\leq \texttt{right}_B, b_i\leq j \leq \texttt{top}_B \}.$ 
\end{lemma}

Taking the block colored in red in \Cref{figure:bss-blocks} (b) as an example, we have $\texttt{left}_{B}=1$, $\texttt{right}_{B}=3$ and $\texttt{top}_{B}=5$. For each column $i \in [1,3]$, we have $b_1=3,b_2=3,b_3=4$. These parameters precisely describe the shape and position of the block.
Furthermore, as shown in \cref{lemma:sameshape}, all blocks from the same equivalence class have the same shape.
Therefore, we only preprocess one corresponding block for each equivalence class.

\begin{lemma}
\label{lemma:sameshape}
Given a string $T$ and its grid, for each equivalence class $C$ and pair of blocks $B_x \in C, B_y \in C$, the following satisfies: 
\begin{itemize}
    \item $\texttt{right}_{B_x} - \texttt{left}_{B_x} = \texttt{right}_{B_y} - \texttt{left}_{B_y}$
    \item $\forall 0 \leq i \leq \texttt{right}_{B_x} - \texttt{left}_{B_x}$, $\texttt{top}_{B_x} - b_{\texttt{left}_{B_x} + i} = \texttt{top}_{B_y} - b_{\texttt{left}_{B_y} + i}$
\end{itemize}

\end{lemma}

\subsection{Relationships between Blocks}
There are further relationships between neighboring blocks. Here, we denote two blocks as neighbors if they share a common edge on any row or column in the grid. If we treat each equivalence class as a node and establish edges between neighboring classes, the resulting graph would become the CDAWG. However, the key discovery in this paper is not just the use of equivalence classes but the grid structure itself and its connection to the suffix trees, which has not been explored in previous works. The grid structure provides a powerful framework for organizing and querying these classes.

Recall the definition of Suffix Trees. Given a text $T$, let $\texttt{PreTree}(T)$ be the suffix tree $\mathcal{T}(T)$, we have the following lemma.

\begin{lemma}
\label{theorem:neighbor2}
    For an equivalence class $C$, we consider any block $B$ of $C$ on the grid. For any column $i$ ($\texttt{left}_B \le i \le \texttt{right}_B$), there exists a node $v_{B, i}$ on $\texttt{PreTree}(T)$ such that $\texttt{str}(v_{B, i}) = \{T[i, j]\mid b_i \le j \le \texttt{top}_B \}$.
\end{lemma}

The notation $v_{B,i}$ will be used throughout the paper. The following corollary follows:


\begin{corollary}
\label{coro:edge2}
Consider two neighboring blocks $B_1$ and $B_2$ that share a common edge in column $i$. Assuming $B_1$ is on the top of $B_2$, we have $\texttt{parent}(v_{B_1, i})=v_{B_2, i}$ on $\texttt{PreTree}(T)$ . 
\end{corollary}

\begin{figure}[t]
    \centering
    \begin{minipage}[t]{.49\linewidth}
    \vspace{0pt}
    \centering
    \includegraphics[width=\linewidth]{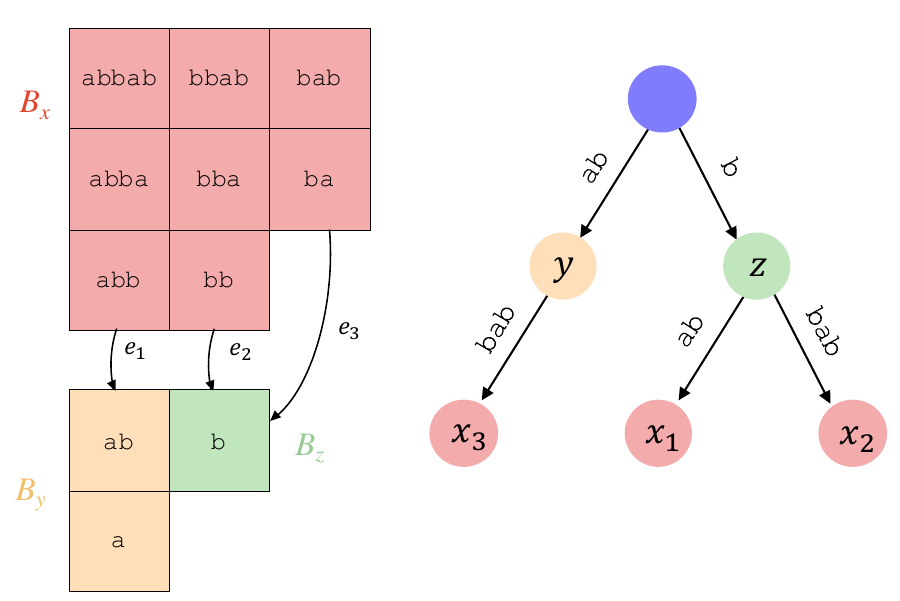}
    \caption{Blocks and $\texttt{PreTree}(T)$}
    \label{figure:bss-edges3}
    \end{minipage}
    \hfill
    \begin{minipage}[t]{.48\linewidth}
        \vspace{0pt}
        \captionof{table}{Nodes and relations in \Cref{figure:bss-edges3}}
        \label{tab:relation}
        \centering
        \resizebox{0.76\linewidth}{!}{%
        \begin{tabular}{|c|c|c|}
            \hline
            node  & column  & $\texttt{str}(\;\cdot\;)$                                       \\ \hline
            $x_1$ & $3_{rd}$ column of $B_x$  & $\{\texttt{abbab}, \texttt{abba}, \texttt{abb}\}$                   \\ \hline
            $x_2$ & $2_{nd}$ column of $B_x$ & $\{\texttt{bbab}, \texttt{bba}, \texttt{bb}\}$    \\ \hline
            $x_3$ & $1_{st}$ column of $B_x$  & $\{\texttt{bab}, \texttt{ba}\}$ \\ \hline
            $y$ & $1_{st}$ row of $B_y$ & $\{\texttt{ab}, \texttt{a}\}$                                 \\ \hline
            $z$   & $1_{st}$ row of $B_z$  & $\{\texttt{b}\}$                                  \\ \hline
        \end{tabular}
        }
        \\[0.3cm]
        \resizebox{0.92\linewidth}{!}{%
        \begin{tabular}{|c|c|c|c|}
        \hline
        & from                & to                  & $\texttt{PreTree}(T)$ \\ \hline
        $e_1$                    & $1_{st}$ row of $B_x$  & $1_{st}$ row of $B_y$  & $x_3$ to $y$                 \\ \hline
        $e_2$                    & $2_{nd}$ row of $B_x$ & $2_{nd}$ row of $B_y$ & $x_2$ to $z$                 \\ \hline
        $e_3$                    & $1_{st}$ row of $B_y$  & $1_{st}$ row of $B_z$  & $x_1$ to $z$                   \\ \hline
        \end{tabular}
        }
        \end{minipage}
\end{figure}

By \cref{coro:edge2}, the relationship between blocks sharing a column indicates an ancestor-descendant relationship on $\texttt{PreTree}(T)$. Furthermore, if they are adjacent, they have a parent relationship. We further map such relationships onto the grid by adding directed edges. In \Cref{figure:bss-edges3}, the blocks $B_x$, $B_y$ and $B_z$ correspond to the nodes $\{x_1,x_2,x_3\}$, $\{y\}$ and $\{z\}$, respectively. The solid black arrows represent the edges in $\texttt{PreTree}(T)$. \cref{tab:relation} shows detailed information about these nodes and edges.

Similarly, we build one other tree $\texttt{SufTree}(T)$ to help formalize the relationship between blocks sharing columns, which maintains substrings with common starting positions. 
Specifically, we build a Suffix Tree of $\rev(T)$. Then, for each node $u$, we reverse every string in $\texttt{str}(u)$.
We denote the tree as $\texttt{SufTree}(T)$. 
The $\texttt{SufTree}(T)$ with $T = \texttt{abbab}$ is shown in \Cref{figure:bss-edges2}. For example, $\texttt{str}(x_2)=\{\texttt{abba},\texttt{bba},\texttt{ba}\}$ (after reversion).

\begin{lemma}
\label{theorem:neighbor1}
    For an equivalence class $C$, we consider any block $B$ of $C$ in the grid. For any row $j$ $(b_{\texttt{left}_B} \le j \le \texttt{top}_B)$, its right boundary $r_j$ is the maximum $i \in [\texttt{left}_B, \texttt{right}_B]$  such that $j \geq b_{i}$. Furthermore, there exists a node $u_{B, j}$ on $\texttt{SufTree}(T)$ such that $\texttt{str}(u_{B, j}) = \{T[i, j]\mid \texttt{left}_B \le i \le r_j \}$.
\end{lemma}

The proof is similar to \cref{theorem:neighbor2}'s. The notation $u_{B,j}$ will also be used throughout the paper. We have the following corollary similar to \cref{coro:edge2}.

\begin{corollary}
\label{coro:edge1}
Consider two neighboring blocks $B_1$ and $B_2$ that share a common edge in row $j$. Assuming $B_1$ is to the left of $B_2$, we have $\texttt{parent}(u_{B_1, j})=u_{B_2, j}$ on $\texttt{SufTree}(T)$ . 
\end{corollary}

By \cref{coro:edge1}, the relationship between blocks sharing a row indicates an ancestor-descendant relationship on $\texttt{SufTree}(T)$. Furthermore, if they are adjacent, they have a parent relationship. We further map such relation onto the grid by adding directed edges, shown in \Cref{figure:bss-edges2}. By \cref{coro:edge1} and \cref{coro:edge2}, we have the following lemma.
\begin{lemma}
\label{lemma:linear-rows}
Given a string $T$ and all distinct blocks $B_1, B_2, \ldots, B_k$ in its grid, the sums of the length of the left boundary and the top boundary of all distinct blocks are both $O(n)$. Formally,
$
\sum_{i=1}^k (\texttt{right}_{B_k} - \texttt{left}_{B_k} + 1) = O(n) \text{ and }
\sum_{i=1}^k (\texttt{top}_{B_k} - b_{\texttt{left}_{B_k}} + 1) = O(n)
$.
\end{lemma}

\noindent
The proofs of \cref{lemma:blockshape,lemma:sameshape,theorem:neighbor2,lemma:linear-rows} can be found in \cref{sec:proof}.

Now, we are ready to formally introduce the \textit{Basic Substring Structure} $\BSS(T)$ of a given text $T$, which consists of the following components:
\begin{itemize}
    \item Two modified suffix trees $\texttt{SufTree}(T)$, $\texttt{PreTree}(T)$ and their mapping edges on $G$.
    \item  $\texttt{left}_B, \texttt{right}_B, \texttt{top}_B$, and $b_{\texttt{left}_B} \leq b_{\texttt{left}_B+1} \leq \ldots \leq b_{\texttt{right}_B}$ for a representative block $B$ in each equivalence class;
    \item A data structure that locates a given substring $T[l,r]$'s equivalence class and its position within the block in $O(1)$ time.
\end{itemize}

By \cref{lemma:linear-rows}, $\BSS(T)$ takes up to 
$O(n)$ memory.

\begin{lemma}
\label{lemma-bss-contruction}
    Given a string $T$, $\BSS(T)$ can be constructed in $O(n)$ time.
\end{lemma}

\begin{proof}
By \cite{crochemore1997compact}, $\texttt{PreTree}(T)$ and $\texttt{SufTree}(T)$ can be constructed in $O(n)$.
By \cref{theorem:neighbor1} and \cref{theorem:neighbor2}, for each node $u$ in $\texttt{PreTree}(T)$ and $\texttt{SufTree}(T)$, all the strings in $\texttt{str}(u)$ belong to a column (row) of a same equivalence class $C$.
Once we can determine the equivalence class that each node's string belongs to, we can compute the shapes of all blocks in the grid $G$, i.e., $\texttt{left}_B, \texttt{right}_B, \texttt{top}_B$, and $b_{\texttt{left}_B} \leq b_{\texttt{left}_B+1} \leq \ldots \leq b_{\texttt{right}_B}$, in $O(n)$ time.

\begin{figure}[t]
    \centering
    \begin{minipage}[t]{.49\linewidth}
    \vspace{0pt}
    \centering
    \includegraphics[width=\linewidth]{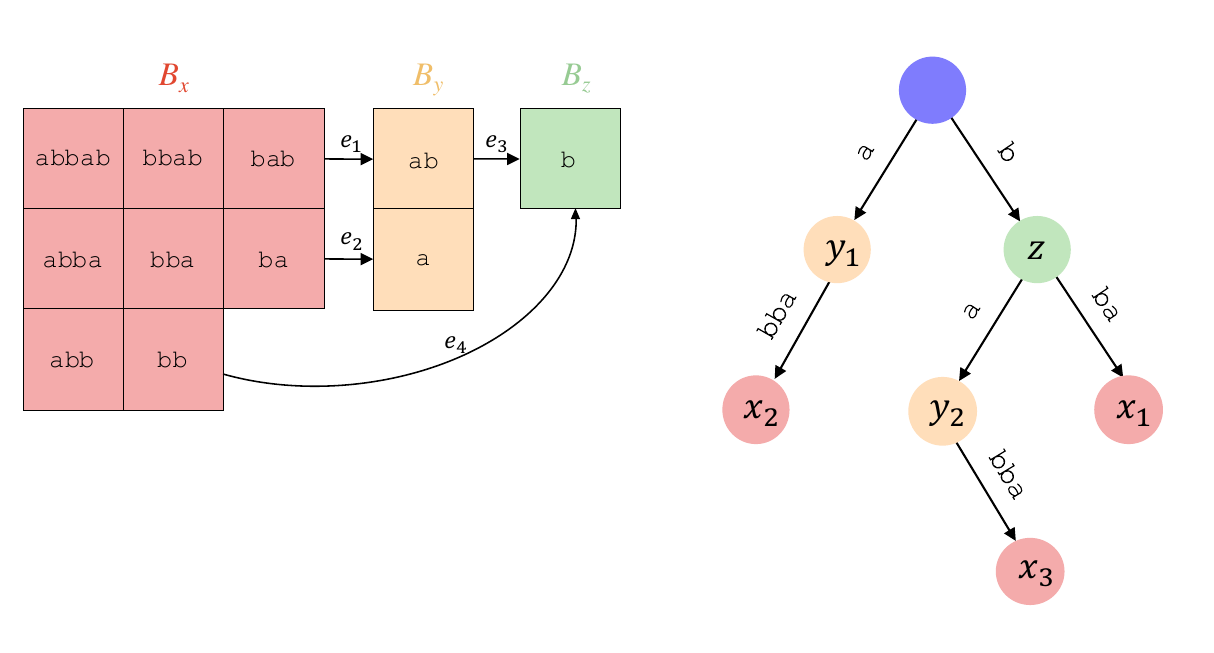}
    \caption{Blocks and \texttt{SufTree}$(T)$}
    \label{figure:bss-edges2}
    \end{minipage}
    \hfill
    \begin{minipage}[t]{.48\linewidth}
        \vspace{0pt}
        \captionof{table}{Nodes and relations in \Cref{figure:bss-edges2}}
        \centering
        \resizebox{0.76\linewidth}{!}{%
            \begin{tabular}{|c|c|c|}
            \hline
            node  & row  & $\texttt{str}(\;\cdot\;)$                                    \\ \hline
            $x_1$ & $1_{st}$ row of $B_x$  & $\{\texttt{abb}, \texttt{bb}\}$                   \\ \hline
            $x_2$ & $2_{nd}$ row of $B_x$ & $\{\texttt{abba}, \texttt{bba}, \texttt{ba}\}$    \\ \hline
            $x_3$ & $3_{rd}$ row of $B_x$  & $\{\texttt{abbab}, \texttt{bbab}, \texttt{bab}\}$ \\ \hline
            $y_1$ & $1_{st}$ row of $B_y$  & $\{\texttt{a}\}$                                  \\ \hline
            $y_2$ & $2_{nd}$ row of $B_y$ & $\{\texttt{ab}\}$                                 \\ \hline
            $z$   & $1_{st}$ row of $B_z$  & $\{\texttt{b}\}$                                  \\ \hline
            \end{tabular}
        }
        \\[0.3cm]
        \resizebox{0.92\linewidth}{!}{%
            \begin{tabular}{|c|c|c|c|}
            \hline
             & from                & to                  & $\texttt{SufTree}(T)$ \\ \hline
            $e_1$                    & $1_{st}$ row of $B_x$  & $1_{st}$ row of $B_y$  & $x_3$ to $y_2$                 \\ \hline
            $e_2$                    & $2_{nd}$ row of $B_x$ & $2_{nd}$ row of $B_y$ & $x_2$ to $y_1$                 \\ \hline
            $e_3$                    & $1_{st}$ row of $B_y$  & $1_{st}$ row of $B_z$  & $y_2$ to $z$                   \\ \hline
            $e_4$                    & $3_{rd}$ row of $B_x$  & $1_{st}$ row of $B_z$  & $x_1$ to $z$                   \\ \hline
            \end{tabular}
        }
        \end{minipage}
\end{figure}

To achieve this, we define the representative string $\texttt{rep}(C)$ as the longest string in each equivalence class $C$, which is unique.
Note that, $T[l,r]$ is a representative string if and only if the following satisfies: \textbf{(1) } $|\texttt{occ}(T[l,r])|\neq |\texttt{occ}(T[l-1,r])| \text{ or } l = 1;$ \textbf{(2) } $|\texttt{occ}(T[l,r])|\neq |\texttt{occ}(T[l,r+1])| \text{ or } r = n.$

This implies $\texttt{rep}(C)$ can only be among the longest strings of $\texttt{SufTree}(T)$'s node ($\texttt{PreTree}(T)$'s node).
Thus, we have
\begin{align*}
\{\texttt{rep}(C) \mid C \in T\text{'s equivalence classes}\} &\subseteq \{\texttt{maxstr}(u)\mid u \in \texttt{SufTree}(T)\} \\
\{\texttt{rep}(C) \mid C \in T\text{'s equivalence classes}\} &\subseteq \{\texttt{maxstr}(u)\mid u \in \texttt{PreTree}(T)\}
\end{align*}

In addition, computing the number of occurrences of a string can be done in $O(1)$ time on the suffix tree with $O(n)$ preprocessing time. \cite{belazzougui2021weighted} shows that it is possible to locate the node corresponding to any substrings $T[l, r]$ on the suffix tree in $O(1)$ time by using $O(n)$ preprocessing time and space. Therefore, we can take $O(n)$ time to check this for all $\texttt{maxstr}(u)$ from $\texttt{SufTree}(T)$ ($\texttt{PreTree}(T)$), and thus locate all representative strings in the suffix trees. 

To infer the equivalence class of each node $u$ in $\texttt{PreTree}(T)$ (or $\texttt{SufTree}(T)$), we first check if $\texttt{maxstr}(u)$ is a representative string. If it is, then $u$ belongs to the equivalence class of $\texttt{maxstr}(u)$. If it is not, we locate $\texttt{maxstr}(u)$ in the opposite suffix tree ($\texttt{SufTree}(T)$ for $\texttt{PreTree}(T)$, and vice versa), denoting it as node $v$. Since $\texttt{maxstr}(v)$ must be a representative string, $u$ then belongs to the equivalence class of $\texttt{maxstr}(v)$.

To implement the data structure that locates a given substring $T[l,r]$, one can first identify its position in $\texttt{PreTree}(T)$, then determine the corresponding equivalence class and the specific row to which it belongs, and ultimately return the desired grid cell.
\end{proof}

\subsection{Optimal Algorithm for Occurrence Counting} \label{sec:count}
After building $\BSS(T)$, we can mark only $O(d)$ points to answer $\textsc{Count}$ query, optimized from the algorithm mentioned in \cref{sec:gridsystem}. Recall the naive algorithm to answer $\textsc{Count}(l, r)$. We have
$\textsc{Count}(l, r) = \sum_{i=l}^r \sum_{j=i}^r \mathds{1} \{T[i, j] \in \mathcal{D}\}$. We use partial sums on BASS to optimize the above algorithm. Define
\[ \texttt{PreCount}(T[i, j]) = \sum_{k=i}^j \mathds{1} \{T[i, k] \in \mathcal{D} \}\text{, } \texttt{SubCount}(T[i, j]) = \sum_{k=i}^j \texttt{PreCount}(T[k, j]).
\]
We can see that $\textsc{Count}(l, r)$ is equal to $\texttt{SubCount}(T[l, r])$. However, storing these values for all substrings $T[i,j]$ is expensive, so we store the following in the data structure:
\begin{enumerate}
    \item $\texttt{SubCount}(T[i,j])$ for strings located at the left border of each distinct block $B$.
    \item Partial sums of $\texttt{PreCount}(\texttt{maxstr}(\texttt{parent}(v_{B,i})))$ for $\texttt{left}_B\leq i\leq \texttt{right}_B$ for each distinct block $B$.
    \item A static 2D range counting data structure built on pattern grid cells in $B$.
\end{enumerate}

Queries can be answered through the following lemma: 
\begin{lemma}\label{lemma:subcount}
    Let $B$ be a block containing the substring $T[l,r]$. Then
    \[
    \begin{aligned}
    \textsc{Count}(l,r)&=\texttt{SubCount}(\texttt{maxstr}(\texttt{parent}(u_{B,r})))\\&\phantom{{}={}}+\sum_{l\leq i\leq r,T[i,r]\in B}\texttt{PreCount}(\texttt{maxstr}(\texttt{parent}(v_{B,i}))) + C,
    \end{aligned}
    \]
    where $u_{B,r}$ is the node of $\texttt{SufTree}(T)$ corresponding to row $r$ of $B$, and $C$ is the answer to a 2D range counting $[l,\infty]\times [-\infty,r]$ only considering the pattern grid cells in the block $B$.
\end{lemma}

The first term is already stored in the data structure, as $\texttt{maxstr}(\texttt{parent}(u_{B,r}))$ is on the left border. The second term can be computed in constant time by the partial sums. The last term is handled by the static 2D range counting data structure.

Now we show how to preprocess these values efficiently. First, locating patterns in BASS costs $O(d)$ time. Then, we adopt the following result from \cite{chan2010counting} to construct the 2D range counting data structure: 

\begin{theorem}{\normalfont\cite{chan2010counting}}
    There is a data structure that can preprocess $n$ points in the plane in $O(n\sqrt{\log n})$ time, using $O(n)$ words of space, and count the number of points dominated by a query point in $O(\log n/ \log \log n)$ time.
\end{theorem}

To preprocess the remaining information, we need to preprocess $\texttt{PreCount}(\texttt{maxstr}(v))$ for all nodes $v\in \texttt{PreTree}(T)$, then do a partial sum, and compute $\texttt{SubCount}(\texttt{maxstr}(u))$ for all node $u\in \texttt{SufTree}(T)$.
First, $\texttt{PreCount}(\texttt{maxstr}(v))$ is the number of patterns that is a prefix of $\texttt{maxstr}(v)$, so it can be computed by getting the number of patterns located on each node of $\texttt{PreTree}(T)$ and summing up these values along the edges between the root and $v$.
Computing $\texttt{SubCount}(\texttt{maxstr}(u))$ is a bit more difficult. We also use the equation of \cref{lemma:subcount}. The first two terms can be computed in $O(1)$ time after taking a partial sum on $\texttt{PreCount}$. For the last term $C$, notice that all $\texttt{maxstr}(u)$ in a block $B$ have the same and smallest $l$-coordinate, $\texttt{left}_B$. Thus, the answer to the 2D range counting query is the number of patterns on or below the line of $u$ in block $B$. Consequently, after preprocessing the number of patterns on each node of $\texttt{SufTree}(T)$, values of $\texttt{SubCount}(\texttt{maxstr}(u))$ in a block $B$ can be computed in time linear to the height of the block. Thus, preprocessing costs $O(n+d\sqrt{\log n})$ time in total.
Putting pieces together, we have the following theorem:

\counttheorem*

Moreover, we can further prove that our algorithm is equivalent to 2D range counting.

\begin{theorem}\label{count3}
    If \textsc{Count}$(l, r)$ can be answered in $O(Q(n,d))$ time after $O(T(n,d))$ time preprocessing, then for every 2D range counting instance with $n$ points whose coordinates are in $\{1,2,\dots,n\}$, there is a data structure that can answer queries in $O(1+Q(2n,n))$ time after $O(n+T(2n,n))$ time preprocessing.
\end{theorem}

On the other hand, the optimal query time complexity for the 2D range counting problem is $O(\frac{\log n}{\log\log n})$ when the data structure has a size of $\tilde{O}(n)$ \cite{Pat07}. Consequently, our query time for \textsc{Count}$(l,r)$ queries is optimal. Moreover, the preprocessing time of $O(d\sqrt{\log n})$ is currently the state-of-the-art in this field.

\noindent
The proofs of \cref{lemma:subcount,count3} can be found in \cref{sec:proof}.

\section{On Querying Distinct Patterns}
\label{sec:countdistinct}
In this section, we discuss the \textsc{CountDistinct} queries, answering the open question from \cite{charalampopoulos2020counting} affirmatively.
To answer \textsc{CountDistinct} queries, we convert them into a data structure problem using BASS and \textsc{Count} queries in \Cref{sec:count}. Our main result is as follows:

\countdistinct*

\subsection{Problem Decomposition}
\label{sec:trans}
Given a text $T$, we consider $\texttt{PreTree}(T)$ of $\BSS(T)$.
For each node $u$ in $\texttt{PreTree}(T)$, define $f_u=\texttt{PreCount}(\texttt{maxstr}(u))$, counting the number of unique patterns in $\mathcal{D}$ that can be recognized by $u$ itself and the ancestors of $u$ in $\texttt{PreTree}(T)$.
We also define a trie $\texttt{Trie}(T[l, r])$ built on all suffixes of $T[l, r]$. For a node $u$ of the trie, let $\texttt{str}'(u)$ be the string formed by the path from the root to $u$. We define $f'_u=\texttt{PreCount}(\texttt{str}'(u))$.
Now, we have the following lemma that calculates $\textsc{CountDistinct}(l, r)$ on $\texttt{Trie}(T[l, r])$:

\begin{lemma}\label{lemma:A1A2}
\begin{align*}
\textsc{CountDistinct}(l, r) &= \underbrace{\sum\limits_{u \text{ is a leaf node of } \texttt{Trie}(T[l, r])} f'_u}_{A_1} \\ 
&- \underbrace{\sum\limits_{u \text{ is a non-leaf node of } \texttt{Trie}(T[l, r])} (\cnt'_u - 1) f'_u}_{A_2} \\
&= A_1 - A_2,
\end{align*}
where $\cnt'_u$ represents the number of children of $u$ in $\texttt{Trie}(T[l, r])$.
\end{lemma}

We have divided the \textsc{CountDistinct} queries into two parts, $A_1$ and $A_2$. Now, we further explain what they are. To assist, we define the following key position:
\begin{definition}[Key Position]
\label{def:pos}
Given a string $T$ and a query $\textsc{CountDistinct}(l, r)$, the key position $\pos$ is defined as the index in $[l,r]$ such that $T[\pos,r]$ is the shortest suffix that appears exactly once in $T[l,r]$.
\end{definition}

With the definition of the key position, we can show that $A_1$ can be computed as follows:

\begin{lemma} \label{lemma:A1}
    $A_1 = \textsc{Count}(l,r)-\textsc{Count}(\pos + 1, r)$.
\end{lemma}

Then, $A_2$ can be defined on $\texttt{PreTree}(T)$ with the following lemma:

\begin{lemma} \label{lemma:A2}
    $A_2 = \sum_{u\in \texttt{PreTree}(T), \cnt_u \geq 1} (\cnt_u-1)f_u$, where $\cnt_u$ represents the number of children of $u$ in $\texttt{PreTree}(T)$ that recognizes at least one substring in $T[l,r]$.
\end{lemma}

Combining \cref{lemma:A1A2} with \cref{lemma:A1} and \ref{lemma:A2}, we get rid of the auxiliary trie and obtain:

\begin{lemma}\label{theo:countdistinct}
$$\textsc{CountDistinct}(l, r) = \textsc{Count}(l,r)-\textsc{Count}(\pos + 1, r)-\sum_{u\in \texttt{PreTree}(T), \cnt_u \geq 1} (\cnt_u-1)f_u,$$ where $\cnt_u$ represents the number of children of $u$ in $\texttt{PreTree}(T)$ that recognize at least one substring in $T[l,r]$.
\end{lemma}

To efficiently answer $\textsc{CountDistinct}(l, r)$, we need: \textbf{(1)} Find the key position $\pos$. \textbf{(2)} Compute $\textsc{Count}(l,r)$ and $\textsc{Count}(\pos + 1, r)$ by the algorithm in \Cref{sec:count}. \textbf{(3)} Compute  $A_2=\sum_{u\in \texttt{PreTree}(T), \cnt_u \geq 1} (\cnt_u-1)f_u$.
In \cref{sec:findkey}, we discuss the solutions to \textbf{(1)}. \textbf{(3)} is discussed in detail in \Cref{sec:solvelct}.

\noindent
The proofs of \cref{lemma:A1A2,lemma:A1,lemma:A2} can be found in \cref{sec:proof}.

\subsection{Finding Key Positions}
\label{sec:findkey}

Recall the definition of key positions: $T[\pos, r]$ occurs exactly once in $T[l,r]$, which means that the previous occurrence of $T[\pos, r]$ has an end position smaller than $l+r-\pos$. Thus, we define $a_0^{(r)}(t)$ to be the end position of the last occurrence of the string $t$ in $T[1,r]$ ($0$ if no occurrence). By definition, $\pos$ is the largest number that has $a_0^{(r-1)}(T[\pos,r])<l+r-\pos$. Consider each node $u$ in $\texttt{SufTree}(T)$. Since all strings in $\texttt{str}(u)$ share the same end positions, these strings have the same value of $a_0^{(r)}(t)$ for every fixed $r$. So we can regard $a_0^{(r)}(t)$ as a variable attached to each node of $\texttt{SufTree}(T)$, and we define $a^{(r)}(u)=a_0^{(r)}(\texttt{maxstr}(u))$.

Now we show how to find the key position for a query \textsc{CountDistinct}$(l,r)$ if $a^{(r)}(u)$ is maintained for all $0\leq r\leq n$ and $u\in \texttt{SufTree}(T)$. Let $x_r$ be the node that $T[1,r]\in \texttt{str}(x_r)$.

\begin{lemma}
\label{lemma:p1-string-to-tree}
Let $y\in \texttt{SufTree}(T)$ be the lowest-depth node on the path from the root to $x_r$ that satisfies $a^{(r-1)}(y)-\texttt{len}(y)+1<l$. We have 
$\pos = \min\{r + l - a^{(r-1)}(y) - 1, r - \texttt{len}(\texttt{parent}(y))\}$, where we define $\texttt{len}(\texttt{parent}(y)) = 0$ when $y$ is the root of $\texttt{SufTree}(T)$.
\end{lemma}

\subsubsection{Data Structure} Now we introduce a data structure that supports finding the node $y$ stated in \cref{lemma:p1-string-to-tree} for a given query range $[l,r]$. We focus on the difference between $a^{(r-1)}$ and $a^{(r)}$. The changes take place in strings having a new occurrence, that is, they are a suffix of $T[1,r]$. We have 
\[
a^{(r)}(u)=\begin{cases}
    r & u\text{ is on the path from the root to } x_r\\
    a^{(r-1)}(u) & \text{otherwise.}
\end{cases}
\]
Then, we define the following events to characterize the difference between $a^{(r-1)}$ and $a^{(r)}$.

\begin{definition}
\label{def:event-seq}
    An event $(i, j, u, v)$ is defined as follows:
    \begin{itemize}
        \item $u$ and $v$ are vertices in $\texttt{SufTree}(T)$ satisfying $u = v$ or $v$ is $u$'s ancestor.
        \item $a^{(i-1)}(x)=j,a^{(i)}(x)=i$ for all vertices $x$ on the path from $u$ to $v$.
    \end{itemize}
\end{definition}

Then, the following lemma shows that all these changes (for $r=1,2,\dots,n$) can be represented by a near-linear number of events. We utilize the \textit{Link-Cut Tree} \cite{sleator1981data} here. 
\begin{lemma}
    \label{lemma:event-seq}
    There exists sets of events $E_1,E_2,\dots,E_n$ satisfying:
    \begin{itemize}
        \item The paths from $u$ to $v$ of all events $(i,j,u,v)\in E_r$ do not intersect and their union form the path from $x_r$ to the root.
        \item $a^{(r)}$ can be obtained from $a^{(r-1)}$ by changing $j$ to $i$ for all nodes on the path from $u$ to $v$ for all events $(i,j,u,v)$ in $E_r$.
        \item $\sum_{i=1}^n |E_i|=O(n\log n)$ and $E_1,E_2,\dots,E_n$ can be constructed in $O(n\log n)$ time.
    \end{itemize}
\end{lemma}

These sets of events give us enough information to recover all the values of $a^{(r)}(u)$. When querying the key position, we only care about the values of $a^{(r-1)}(u)$ for $u$ on the path from the root to $x_r$, and these values are stored by those $j$ in events $(i,j,u,v)\in E_r$.
We now show how to find the key position by \cref{lemma:p1-string-to-tree} when given $E_r$. The procedure has two steps. The first step is to find the node $y$ with the smallest depth that satisfies $a^{(r-1)}(y)-\texttt{len}(y)+1<l$. Note that the value of $a^{(r-1)}(y)$ decreases on the path from the root to $x_r$ because every time the longer substring occurs, the shorter substring also occurs. Thus, $a^{(r-1)}(y)-\texttt{len}(y)+1$ also decreases on the path from the root to $x_r$. Consequently, we can use a binary search on $E_r$ to locate $y$. The second step is to find $\pos$ when given $y$, which is a constant-time evaluation.
Overall, combined with \cref{lemma:p1-string-to-tree}, we have the following lemma:

\begin{lemma}
    \label{lemma:finding-key-positions}
    There exists a data structure of $O(n \log n)$ size that finds the key position $\pos$ of $\textsc{CountDistinct}(l, r)$ in $O(\log n)$ time with a precomputation of $O(n\log n)$ time.
\end{lemma}

\noindent
The proofs of \cref{lemma:p1-string-to-tree,lemma:event-seq} can be found in \cref{sec:proof}.

\subsection{\texorpdfstring{Computing $A_2$}{Computing A2}}
\label{sec:solvelct}

Recall that $A_2=\sum_{u\in \texttt{PreTree}(T), \cnt_u>1} (\cnt_u-1)f_u$, since $f_u$ is fixed by the dictionary $\mathcal{D}$ and does not depend on the query range $[l,r]$, the challenge of computing $A_2$ is to maintain the values of $\cnt_{u}$ for all nodes in $\texttt{PreTree}(T)$ and different query ranges.
Similarly, we make use of the sets of events $E_1, E_2,\dots, E_n$ defined in \cref{lemma:event-seq}, but on $\texttt{PreTree}(T)$. Specifically, since the substrings in $\texttt{str}(v)$ for $v\in \texttt{PreTree}(T)$ have the same starting positions, we define $b^{(l)}(v)$ to be the start position of the first occurrence of $\texttt{maxstr}(v)$ in $T[l,n]$ ($n+1$ if no occurrence). And, the difference between $b^{(l+1)}$ and $b^{(l)}$ is described by the event set $F_l$. Formally, we define the events again:
\begin{definition}
    An event $(i,j,u,v)$ is defined as follows:
    \begin{itemize}
        \item $u$ and $v$ are vertices in $\texttt{PreTree}(T)$ satisfying $u=v$ or $v$ is $u$'s ancestor.
        \item $b^{(i+1)}(x)=j,b^{(i)}(x)=i$ for all vertices $x$ on the path from $u$ to $v$.
    \end{itemize}
\end{definition}
And after defining $x_l$ to be the node such that $T[l,n]\in \texttt{str}(x_l)$, we present the following:
\begin{lemma}
    There exist sets of events $F_1,F_2,\dots,F_n$ satisfying:
    \begin{itemize}
    \item The paths from $u$ to $v$ of all events $(i,j,u,v)\in F_l$ do not intersect and their union form the path from $x_l$ to the root.
    \item $b^{(l)}$ can be obtained from $b^{(l+1)}$ by changing $j$ to $i$ for all nodes on the path from $u$ to $v$ for all events $(i,j,u,v)$ in $F_l$.
    \item $\sum_{i=1}^n |F_i|=O(n\log n)$ and $F_1,F_2,\dots,F_n$ can be constructed in $O(n\log n)$ time.
    \end{itemize}
\end{lemma}

Let $A_2(l,r)$ represent the value of $A_2$ when the query range is $[l,r]$. Assuming that we already know the value of $A_2(l+1, r)$, we consider how to derive the value of $A_2(l, r)$. The difference comes from the substrings that occur in $T[l,r]$ but not in $T[l+1,r]$, changing the value of $\cnt$.
Assume that we have already maintained the values of $\cnt$ for every node in $\texttt{PreTree}(T)$ when the query range is $[l+1,r]$. For an event $(i,j,u,v)\in F_l$, we consider how it changes the values of $\cnt$.
During the event, we will change the value of \(b^{(l+1)}(x)\) from \(j\) to \(i\) for all vertices $x$ on the path from \(u\) to \(v\), meaning that some of the vertices $x$ may recognize substrings in $T[l,r]$ while they recognize no substrings in $T[l+1,r]$. This possibly changes the values of $\cnt_x$ for the vertices $x$ in the path from $\texttt{parent}(u)$ to $\texttt{parent}(v)$.

\begin{lemma}\label{lemma:increment-condition}
    For each vertex $x$ on the path from $u$ to $v$, $\cnt_{\texttt{parent}(x)}$ increases by one if and only if $r\in [i + \texttt{len}(\texttt{parent}(x)), j + \texttt{len}(\texttt{parent}(x)))$.
\end{lemma}

Now we know that when moving the query range from $[l+1,r]$ to $[l,r]$, some of the values of $\cnt$ increase by $1$. Recall that $A_2=\sum_{u\in \texttt{PreTree}(T),\cnt_u> 1} (\cnt_u-1)f_u$. Thus, $A_2(l,r)$ is equal to $A_2(l+1,r)$ plus the sum of $f_{\texttt{parent}(x)}$ for those $x$ whose $\cnt_{\texttt{parent}(x)}$ increased and with $\cnt_{\texttt{parent}(x)}\geq 2$.
By the following lemma, we show that, for a single event, although multiple $\cnt_{\texttt{parent}(x)}$ increased, only one $\cnt_{\texttt{parent}(x)}$ can have $\cnt_{\texttt{parent}(x)}\geq 2$.
\begin{lemma}
\label{lemma:one-in-A2}
Given an event $(i, j, u, v)\in F_l$, let $x, y$ be any pair of vertices on the path from $u$ to $v$ that satisfies $y$ is the parent of $x$.
For all $r \in [i + \texttt{len}(y), j + \texttt{len}(y))$, $\cnt_{y} = 1$.
\end{lemma}

Thus, the only possible $\cnt_{\texttt{parent}(x)}\geq 2$ is at the top of the path, the node $\texttt{parent}(v)$. The following lemma describes when $\cnt_{\texttt{parent}(v)}\geq 2$ is true:

\begin{lemma}
\label{lemma:cnt_more_two_hold}
    Given an event $(i,j_0,u_0,v_0)\in F_l$ and another event $(i,j_1,u_1,v_1)\in F_l$ where $u_1=\texttt{parent}(v_0)$, $\cnt_{u_1}$ increased to $\cnt_{u_1}\geq 2$ if and only if \(r\in [j_1+\texttt{len}(u_1), j_0+\texttt{len}(u_1))\).
\end{lemma}

\subsubsection{Final Construction} Now we conclude the formula for $A_2$. For $l=n,n-1,\dots,1$, we perform the following:
\begin{enumerate}
    \item Sort events in $F_l$ by the order of the paths from $x_l$ to the root.
    \item Let $A_2(l,r)\leftarrow A_2(l+1,r)$ for all $l+1\leq r\leq n$ and $A_2(l,l)\leftarrow 0$.
    \item For all adjacent events $(i,j_0,u_0,v_0)\in F_l$ and $(i,j_1,u_1,v_1)\in F_l$ where $u_1=\texttt{parent}(v_0)$, let $A_2(l,r)\leftarrow A_2(l,r)+f_{u_1}$ for all $r\in [j_1+\texttt{len}(u_1), j_0+\texttt{len}(u_1))$.
\end{enumerate}

By previous lemmas, we prove that the correct value of $A_2(l,r)$ is obtained after this procedure.
Since the total number of events is $O(n\log n)$, the above process can be maintained by $O(n\log n)$ range additions with persistent data structures \cite{driscoll1986making}: With a persistent segment tree, we answer each $A_2$ query in $O(\log n)$ time. We have \Cref{lemma:computingA2} for computing $A_2$.

\begin{lemma}
    \label{lemma:computingA2}
    There exists a data structure of $O(n \log^2 n)$ size that computes the value of $A_2$ of $\textsc{CountDistinct}(l, r)$ in $O(\log n)$ time with a precomputation of $O(n\log^2 n)$ time.
\end{lemma}

\noindent
The proofs of \cref{lemma:increment-condition,lemma:one-in-A2,lemma:cnt_more_two_hold} can be found in \cref{sec:proof}.

\subsection{Conclusion}
Applying \Cref{lemma:finding-key-positions} and \Cref{lemma:computingA2} to the algorithm scheme discussed in \Cref{sec:trans}, we obtain the results of \Cref{CD-res}.
In the study by Charalampopoulos et al. \cite{charalampopoulos2020counting}, the authors examined the specific scenario where the dictionary comprises square strings. They demonstrated an algorithm that achieves a query time of $O(\log n)$ with space and preprocessing time complexities of $O(n\log^2 n)$. Our algorithm matches this.
Furthermore, when the dictionary consists of all the distinct substrings of $T$, computing $A_1$ is straightforward since the result for \textsc{Count}$(l,r)$ queries is given by $\frac{(r-l+1)(r-l+2)}2$. In addition, $A_2$ can be obtained by the algorithm above since $f_u=\texttt{len}(u)$ is easy to compute. Consequently, \textsc{CountDistinct}$(l,r)$ queries can be resolved in $O(\log n)$ time, utilizing $O(n\log^2 n)$ space and preprocessing time.

%% file: more.tex
\section{More Applications}
\label{sec:more}

\subsection{\textsc{Exists} and \textsc{Report}}
\label{sec:e-r}
There are two more fundamental types of queries under the internal setting: \textsc{Exists} and \textsc{Report}. While previous results already approach nearly optimal efficiency, we provide alternative algorithms with our proposed BASS.

\subsubsection{\textsc{Exists}}
\label{sec:exists}

First, we consider \textsc{Exists}$(l,r)$ queries, determining if there exists a pattern from $\mathcal{D}$ that occurs in $T[l,r]$. Our algorithm performance is formally stated as follows:

\begin{restatable}[Exists]{theorem}{exists}
\label{theo:exist}
\textsc{Exists}$(l,r)$ can be answered in $O(1)$ time with a data structure of size $O(n+d)$ that can be constructed in $O(n+d)$ time.
\end{restatable}

Our algorithm for \textsc{Exists}$(l,r)$ queries works in a similar fashion to \textsc{Count}. Recall that in \Cref{sec:count}, we can compute
\begin{align*}
    \sum_{i = l}^r \sum_{j = i}^r \mathds{1} \{T[i, j] \in \mathcal{D} \land T[i, j] \notin B\}
\end{align*}

in $O(1)$ time after $O(n + d)$ preprocessing, where $B$ is the block to which $T[l, r]$ belongs. If this value is nonzero, the algorithm returns with positive validation; otherwise, $\textsc{Exists}(i, j) = \texttt{True}$ if and only if there exists $[i, j] \subseteq [l, r]$ such that $T[i, j] \in \mathcal{D}$ and $T[i, j] \in B$. It suffices to show how the existence of such $[i, j]$ can be decided in $O(1)$ time given $[l, r]$.

To solve this, we consider each equivalence class in $\BSS(T)$ independently.
Similar to \cref{sec:count}, we only need to consider an arbitrary block of this class.
By regarding the pattern cells in the block as $m$ normalized points in the 2D plane, this problem is thus equivalent to the following: 
\begin{problem}
Given $m$ points $(x_i, y_i)$ on the 2D plane ($x_i, y_i \in [1, m]$), check whether the region $[l, +\infty) \times (-\infty, r]$ contains any point. 
\end{problem}
By precomputing the values of $\min_{(x_i, y_i): x_i \geq v} y_i, \forall v \in [1, m]$ in $O(m)$ time, each query to this problem can be answered in $O(1)$ time.
Recall that the sum of $m$ for all equivalence classes is $O(d)$. We thus prove \cref{theo:exist}.

\subsubsection{\textsc{Report}}
\label{sec:report}

Then, we answer \textsc{Report}$(l,r)$ queries, which reports all occurrences of all patterns of $\mathcal{D}$ in $T[l,r]$. Our algorithm performance is formally stated as follows:

\begin{restatable}[Report]{theorem}{report}
\label{theo:report}
\textsc{Report}$(l,r)$ can be answered in $O(1+x)$ time with a data structure of size $O(n+d)$ that can be constructed in $O(n+d)$ time, where $x$ denotes the length of output.
\end{restatable}

Our algorithm for \textsc{Report} consists of two parts: 
\begin{enumerate}
    \item We compute a set $\mathcal{R}$ consisting of all indices $j \in [l, r]$ such that there exists a pattern in $T[l, r]$ with the end position $j$.
    \item For each $j$ in $\mathcal{R}$, we report a set $\mathcal{L}(j)$ of the start positions $i$ such that $T[i, j]$ is a pattern.
\end{enumerate}

Our goal is to finish the first part in $O(|\mathcal{R}|)$ time and the second part in $O(|\mathcal{R}|+\sum_{j\in \mathcal{R}} |\mathcal{L}(j)|)$ time. In this way, the total query time will be $O(x)$, and \cref{theo:report} is achieved.

\paragraph{Reporting $\mathcal{L}(j)$. }
We first consider the second part.
Similar to \cref{sec:count}, for each equivalence class $C$ in $\BSS(T)$, it suffices to consider an arbitrary block $B$. 
For each block $B$ considered, we precompute the following values for each row $j$ of $B$: \textbf{(1)} all pattern grid cells $T[i, j] \in \mathcal{D}$ in row $j$ of $B$ in the increasing order of $i$ and \textbf{(2)} the leftmost pattern grid cell $T[i, j] \in \mathcal{D} \setminus B$ that is in row $j$ and to the right of $B$ (\cref{coro:edge1}).

By \cref{lemma:linear-rows}, the total number of rows over all distinct blocks is $O(n)$, so we can compute all the above information in $O(n + d)$ time.
With these precomputed values, we can report the start positions in $\mathcal{L}(j)$ by first locating the block of $T[l, j]$ and then reporting the pattern grid cells in row $l$ from left to right, within time $O(1)$ for each pattern.
This exactly gives the $O(|\mathcal{R}| + \sum_{j \in \mathcal{R}} |\mathcal{L}(j)|)$ time complexity in total for the second part.
 


\paragraph{Finding $\mathcal{R}$. }
Now, we focus on how to find $\mathcal{R}$ efficiently. 
Recall that for each row $j$, $j \in \mathcal{R}$ if and only if there exists a start position $i (l \leq i \leq j)$ such that $T[i, j]$ is a pattern.
To check this in $\BSS(T)$, we may first locate the cell $T[l, j]$, and then see whether there exists a pattern grid cell $(i, j)$ on the right of $(l, j)$.
Similar to the process of finding $\mathcal{L}(j)$s, for a fixed substring $T[l, j_1]$ that has a pattern as its suffix, we can maintain the preceding row $j_2 < j_1$ in the column $l$ such that $T[l,j_2]$ also has a pattern as its suffix.
However, this might take $\Theta(n^2)$ time and space.

To speed up the above process, we record the preceding equivalence class instead of the row.
For each equivalence class $C$, we still consider only one of its blocks $B$ similar to \cref{sec:count}.
We only maintain the ``predecessor'' in a different block for each column $i$ in $B$, which only takes $O(n)$ space by \cref{lemma:linear-rows}.
Formally, given a block $B$ and a column $i$ in $B$, we will maintain the top row $j$ satisfying \textbf{(1)} $j$ is in a block under $B$ and \textbf{(2)} there exists a $k \in [i, j]$ such that $T[k, j] \in \mathcal{D}$. Dynamic programming on the grid can compute this in $O(n + d)$ time.

With this information, we can find the set of blocks that contain at least one row in $\mathcal{R}$. In addition, each row in $\mathcal{R}$ occurs in exactly one block of this set.
This takes $O(|\mathcal{R}|)$ time by first locating the block that $T[l, r]$ belongs to and then iterating rows with the fixed column $l$. 
By doing so, it remains to find the rows $r$ that belong to $\mathcal{R}$ for each of these blocks.

By \cref{lemma:linear-rows}, given a block $B$, we may maintain the rightmost pattern grid cell $T[i, j] \in \mathcal{D}$ for each row $j\ (l \leq j \leq r)$ in $B$ in $O(n+d)$ time.
Denote this by $h_j$, so that row $j$ is in $\mathcal{R}$ if and only if $l \leq h_j$. Now, it suffices to compute the set $S_B$ of indices $j \in [b_l, \min(\texttt{top}_B, r)]$ that satisfies $h_j \geq l$ in $O(|S_B|)$ time. If we view $(j, h_j)$ as points of a 2D plane, this is exactly the $3$-sided range reporting problem. This is studied in~\cite{3sidereport}, and is efficiently solved by \textit{Range Minimum Query (RMQ)} algorithms~\cite{fischer2006theoretical} with the desired performance.

\subsection{\textsc{ReportDistinct}}
\label{sec:reportdistinct}
Compared to the \textsc{Report} query that finds all occurrences in a substring, \textsc{ReportDistinct} queries only report every pattern once. In \cite{charalampopoulos2021internal}, it achieves $O(\log n + x)$ query time with $O(n \log n + d)$ preprocessing, where $x$ denotes the length of the output. With BASS, we improve the query time to $O(1+x)$. Both previous works and our algorithm take $O(n + d)$ memory. The formal theorem is as follows:

\reportdistinct*
Let $D$ denote the set of patterns that occur in $T[l,r]$. The key idea is to find the patterns in $D$ that are not suffixes of any longer pattern in $D$. We denote this set as $S$. With $S$, we can locate and report other patterns in $\texttt{SufTree}(T)$ in $O(1 + x)$ time by the following lemma.
\begin{lemma}
    \textsc{ReportDistinct}$(l,r)$ queries can be answered in $O(1+x)$ time using $S$, where $x=|D|$.
\end{lemma}

\begin{proof}
First, we directly report all the patterns in $S$.
For each $t\in D\setminus S$, let $v$ be the node in $\texttt{SufTree}(T)$ recognizing $t$. Since $t$ is a suffix of some $t'\in S$, $v$ must be an ancestor of the node $u$ recognizing $t'$. Furthermore, for any node $v$ that is an ancestor of a node $u$ recognizing a substring $t'\in S$, all patterns in $\texttt{str}(v)$ should be reported.

Based on that, starting from the nodes recognizing substrings in $S$, we can iteratively "lift" each node to its deepest ancestor containing a non-reported pattern, until no node can be lifted.
This process ensures that every pattern in $D$ is reported exactly once, achieving $O(1 + x)$ time complexity.
\end{proof}

We determine $S$ by computing the following two subsets of $D$.
For each $i\in [l, r]$, let $p_i$ be the longest pattern suffix of $T[l, i]$, recognized by node $u_i\in \texttt{SufTree}(T)$. Let $v_i$ denote the node recognizing $T[l, i]$. Then, the first subset is:
\[
S_1 = \{ p_i \mid u_i = v_i, i\in [l,r]\}.
\]
Let $q_i$ be the longest pattern suffix of $T[l, i]$ recognized at an ancestor of $v_i$. Then, we define the second subset as:
\[
S_2 = \{ q_i \mid q_i \text{ occurs exactly once in } T[l, i], i\in [l,r]\}.
\]
It suffices to find $S_1$ and $S_2$ efficiently due to the following lemma:
\begin{lemma}
    $S\subseteq S_1\cup S_2$.
\end{lemma}
\begin{proof}
    We first define $S_2'=\{q_i\mid i\in [l,r]\}$. Then, \[S\subseteq \{p_i\mid i\in [l,r]\}= \{p_i\mid u_i=v_i,i\in [l,r]\}\cup \{p_i\mid u_i\neq v_i,i\in [l,r]\}\subseteq S_1\cup S_2'.\]
    We have $S_2\subseteq S_2'$, thus it suffices to prove that $(S_2'\setminus S_2)\cap S=\emptyset$. Consider $q_i\in S_2'\setminus S_2$. By definition, $q_i$ occurs at least twice in $T[l, i]$. Let its first occurrence in $T[l, i]$ correspond to the suffix of $T[l, i']$, where $l\leq i'<i$. This implies that $q_i$ is either equals $q_{i'}$ or is a proper suffix of $q_{i'}$.
    \begin{itemize}
        \item Case 1: If $q_i=q_{i'}$, then $q_{i'}$ occurs exactly once in $T[l, i']$, forcing $q_i\in S_2$. This contradicts $q_i\in S_2'\setminus S_2$.
        \item Case 2: If $q_i$ is a proper suffix of $q_{i'}$, then $q_i\notin S$, as elements of $S$ cannot be proper suffixes of other patterns.
    \end{itemize}
\end{proof}

\paragraph{Finding $S_1$}
Let $\mathcal{Q}=\{i\mid u_i=v_i, i\in [l,r]\}$. $\mathcal{Q}$ resembles the set $\mathcal{R}$ in \textsc{Report}$(l,r)$ queries, but $\mathcal{Q}$ includes indices $i$ only if the pattern suffix of $T[l, i]$ is in the same block as $T[l, i]$ in BASS. Thus, by precomputing the rightmost occurrence of patterns in each row of every block, $\mathcal{Q}$ can be efficiently determined using a $3$-sided range reporting algorithm \cite{3sidereport}.

Once $Q$ is identified, we compute $S_1'$, an extension of $S_1$, as follows: $S_1'$ contains all patterns that are suffixes of $T[l,i]$ and lie in the same block as $T[l,i]$ for all $i\in \mathcal{Q}$. We have $S_1\subseteq S_1'$, and since each row in the BASS corresponds to distinct strings, patterns in $S'_1$ are unique. To construct $S'_1$, we iterate over patterns in row $i$ (for each $i\in \mathcal{Q}$) from right to left, terminating when a pattern is no longer a suffix of $T[l, i]$. This process runs in $O(|S'_1|)$ time, ensuring the total cost remains within $O(1+x)$, where $x$ is the output size.

\paragraph{Finding $S_2$} For each node $u$ in $\texttt{SufTree}(T)$, we define $z_u$ as the longest pattern that is a suffix of $\texttt{maxstr}(\texttt{parent}(u))$ ($\epsilon$ if not exist). Furthermore, we define $w_u$ as the node in $\texttt{SufTree}(T)$ that recognizes $z_u$. Since $q_i$ is not recognized by the same node as $T[l, i]$, $q_i$ depends only on $v_i$---the node that recognizes $T[l, i]$, and satisfies $q_i=z_{v_i}$. Despite that the indices $l, i$ may be different, whether $z_{v_i}$ appears in $T[l, i]$ once only depends on the string $T[l, i]$ itself. And, for strings recognized by the same node $u\in \texttt{SufTree}(T)$, the longer the string, the higher the chance that $z_{v_i}$ has another occurrence. Thus, for each node $u\in \texttt{SufTree}(T)$, there is a boundary in its BASS row, where only the substrings on its right contain $z_u$ exactly once.

To determine the boundary, we use the array $a^{(r)}(\cdot)$ from \cref{sec:findkey}. Specifically, $z_{u}$ appears once in a string $T[j,k]\in \texttt{str}(u)$ if and only if $a^{(k-1)}(w_{u})-|z_{u}|+1<j$. This inequality checks whether the previous occurrence of $z_{u}$ starts before $j$. We can check for an arbitrary occurrence of strings in $\texttt{str}(u)$. The value of $a^{(k-1)}(w_{u})$ can be retrieved from the event set $E_{k-1}$ from \cref{lemma:event-seq}, yielding an $O(n\log n)$ preprocessing time.

Finally, indices $i$ contributing to $S_2$ correspond to the positions where $T[l, i]$ lies to the right of the precomputed boundary for that row. Using 3-sided range reporting (similar to $\mathcal{R}$ in \textsc{Report}$(l,r)$ queries and $\mathcal{Q}$ in the previous $S_1$ part), $S_2$ is determined in $O(|S_2|)$ time.

\begin{remark}
By maintaining only the most recent event set $E_r$ during preprocessing (instead of all historical sets) and immediately dealing with the correlated queries, the memory usage can be further improved to $O(n)$.
\end{remark}

\subsection{Range Longest Common Substring}

Consider the following internal query problem related to the longest common substrings (LCS).
\begin{problem}
    Given two strings $S$ and $T$, each query \textsc{LCS}$(l,r)$ requires to compute the longest common substring between $S$ and $T[l, r]$.
\end{problem} 

By using BASS technique, we have the following result.

\begin{restatable}[Range LCS]{theorem}{rlcstheorem}
\label{theo:rlcs}
    $\textsc{LCS}(l, r)$ can be answered in $O(1)$ time with a data structure of $O(|S|+|T|)$ size that can be constructed in $O(|S|+|T|)$ time.
\end{restatable}

This improves the previous best result of $O(\log n)$ query time \cite{amir2020dynamic} and achieves optimal performance.

We start presenting our data structure by introducing an auxiliary string $A$. Let $A = Sc_1Sc_2T$ be a string formed by concatenating $S, c_1, S, c_2, T$, where $c_1, c_2$ are distinct characters that do not exist in either $S$ or $T$. We construct $\BSS(A)$. Given that $c_1$ (or $c_2$) occurs only once in $A$, every string containing $c_1$ or $c_2$ should belong to the same block as the entire string $A$. Consequently, every substring of $S$, which occurs in $A$ at least twice, belongs to a block different from $A$.
Equivalently, we have the following lemma:
\begin{lemma}
Given an equivalence class $C$ of $\BSS(A)$, either all strings $t \in C$ occur in $S$, or none of the strings $t \in C$ occur in $S$.
\end{lemma}

Now, we have the following, where $B$ is the block that $A[l, r]$ belongs to:
\[
\begin{aligned}
    \textsc{LCS}(l,r) &=\begin{cases}
        \ \ \ \ \ \ \ r-l+1 & A[l,r] \text{ occurs in } S\\
        \underset{\substack{l \leq a \leq b \leq r \\ A[a, b] \in S \text{ and }  A[a, b] \notin B}}{\max} \{b - a + 1\} & \text{otherwise}
    \end{cases}
\end{aligned}
\]

One observation is that the optimal $(a, b)$ above must be on some blocks' top or left border.  During the preprocessing phase, we store the answer $\mathrm{LCS}(l, r)$ for every $(l, r)$ that is on the border of a block. Since we only need to store the answer for one arbitrary block for each equivalence class, the total information we need is $O(|A|)$.

What remains is that given the information of the blocks to the down or right of $B$, after some preprocessing within time linear in the perimeter of $B$, we need to compute $\textsc{LCS}(l,r)$ in $O(1)$ time. This would give a unified approach for answering queries in constant time, as well as precomputing the required information in linear time.

The answers on the bottom or right border are easy to retrieve. And, for a substring $t$ on the top or left border of $B$, the answer is the maximum of the answer of the substrings that occurs in $t$ and on the bottom or right border of $B$. Since the block is staircase-shaped by \cref{lemma:blockshape}, if we regard all substrings on the bottom or right border as a sequence, we only need to take the maximum of the answer inside an interval.

As we know that range maximum queries can be answered in constant time by a $O(n)$ time constructible data structure, our algorithm can be done in linear time.
When query, similarly, we only need to answer a range maximum query within that block, costing only constant time.

In \cite{amir2020dynamic}, another problem is also proposed: Given two strings, $S$ and $T$, of lengths up to $n$, the task is to construct a data structure that computes the LCS between any prefix or suffix of S and any prefix or suffix of T. By employing BASS, we can develop a data structure of size $O(n\log\log n)$, which answers such queries in $O(\log\log n)$, outperforming the previous best-known result. We believe BASS offers deeper insight into the relationships among substrings and will lead to further applications.

%% file: proof.tex
\section{Omitted Proof}
\label{sec:proof}

The following proves there is only one longest substring $s$ for any substring $t \in T$, which indicates that the extension is well-defined.
\begin{proof}[Proof. (Uniqueness of maximal extensions)]


We will prove that there is a unique maximal string $s$ for a fixed substring $t \in T$, such that $s$ contains $t$ and $\occ(s) = \occ(t)$. Here, by maximal, we mean that for any character $a$, neither $\occ(as)$ nor $\occ(sa)$ is equal to $\occ(t)$.

Suppose on the contrary that there exist two different substrings $s_1 = a_1tb_1, s_2 = a_2tb_2$ as maximal extensions of $t$. Without loss of generality, assume that $a_2$ is a suffix of $a_1$ and $b_1$ is a prefix of $b_2$. Then, as $\occ(s_1) = \occ(s_2) = \occ(t)$, every occurence of $t$ should have $a_1$ to its left and $b_2$ to its right. Consequently, $\occ(a_1tb_2) = \occ(t)$, where $s_1, s_2$ are both substrings of $a_1tb_2$, contradicting with the assumption that $s_1, s_2$ are maximal extensions.

Notably, this proof also implies that $\ext(s)$ contains $s$ as its substring exactly once.

\end{proof}

\begin{proof}[Proof of \cref{lemma:blockshape}]

Obviously, if $(l'', r'')$ and $(l, r)$ are in the same block $B$, where $l'' \leq l \leq r \leq r''$, then $(l', r')$ should belong to $B$ for every $l' \in [l'', l], r' \in [r, r'']$.

It remains to prove that every block contains its top-left corner. Formally, for any $1 \leq l' < l \leq r < r' \leq n$, it suffices to prove that if $(l', r), (l, r), (l, r')$ are in the same block $B$, then $(l', r')$ also belongs to $B$. This is a corollary of the uniqueness of maximal extensions: the extension $t = \ext(l', r) = \ext(l, r) = \ext(l, r')$ must contain both $T[l', r]$ and $T[l, r']$, and therefore $t$ contains $T[l', r']$. Note that to show $\ext(l', r) = \ext(l, r)$, we use the fact that $\ext(l, r)$ is not only the longest extension but also the unique maximal one.



\end{proof}

\begin{proof}[Proof of \Cref{theorem:neighbor2}]
Considering any node $u$ in $\texttt{PreTree}(T)$, all strings in $\texttt{str}(u)$ share the same start positions of their occurrences in $T$, which means these strings belong to the same equivalence class $C$ and exactly occupy an entire column in each block of $C$.
Therefore, given a column $i$ in a block $B$, the corresponding node $v_{B, i}$ in $\texttt{PreTree}(T)$ is the node that can recognize $T[i,\texttt{top}_B]$.

\end{proof}

\begin{proof}[Proof of \cref{lemma:sameshape}]

Let $(l_x, r_x)$ be the top-left corner of $B_x$ and $t_x = T[l_x, r_x]$. Then, for every $i, j \in [1, |t_x|]$, $(l_x + i - 1, l_x + j - 1)$ belongs to $B_x$ if and only if $\occ(t_x[i, j]) = \occ(t_x)$. Therefore, the shape of $B_x$ depends only on $t_x$.

Since $B_x$ and $B_y$ are in the same equivalence class, we have $t_x = t_y$ where $t_y$ is defined similarly. Consequently, $B_x$ and $B_y$ are isomorphic up to translations, which is basically what the lemma says.

\end{proof}

\begin{proof}[Proof of \Cref{lemma:linear-rows}]
    The total length of the left boundary of all distinct blocks is the number of nodes in $\texttt{SufTree}(T)$, and the total length of the top boundary of all distinct blocks is the number of nodes in $\texttt{PreTree}(T)$. Since the number of nodes in a suffix tree is $O(|T|)$, we have the desired result.
\end{proof}

\begin{proof}[Proof of \Cref{lemma:subcount}]
    Recalling the relationship between neighboring blocks sharing a row in \cref{coro:edge1}, we have
    $$\begin{aligned}
        \texttt{SubCount}(T[l,r])&=\texttt{SubCount}(\texttt{maxstr}(\texttt{parent}(u_{B,r}))) + \sum_{l\leq i\leq r,T[i,r]\in B} \texttt{PreCount}(T[i,r])\\
        &=\texttt{SubCount}(\texttt{maxstr}(\texttt{parent}(u_{B,r})))\\
        &\phantom{{}={}} + \sum_{l\leq i\leq r,T[i,r]\in B} \texttt{PreCount}(\texttt{maxstr}(\texttt{parent}(v_{B,i})))\\
        &\phantom{{}={}}+\sum_{l\leq i\leq r,T[i,r]\in B}\sum_{i\leq j\leq r,T[i,j]\in B} \mathds{1}\{T[i,j]\in \mathcal{D}\}\\
        &=\texttt{SubCount}(\texttt{maxstr}(\texttt{parent}(u_{B,r})))\\
        &\phantom{{}={}}+\sum_{l\leq i\leq r,T[i,r]\in B}\texttt{PreCount}(\texttt{maxstr}(\texttt{parent}(v_{B,i}))) + C
    \end{aligned}$$
\end{proof}

\begin{proof}[Proof of \Cref{count3}]
    Assume the coordinates are $(l_1,r_1),(l_2,r_2),\dots,(l_n,r_n)$. We construct text $T$ of length $2n$ with $2n$ distinct characters. Let the dictionary be $\mathcal{D}=\{T[l_1,r_1+n],T[l_2,r_2+n],\dots,T[l_n,r_n+n]\}$. Then the answer to a query \textsc{Count}$(l, r)$ is the number of $1\leq i\leq n$ that satisfies $l\leq l_i, r\geq r_i+n$. So, the answer to the 2D range counting query $[x,+\infty]\times [-\infty,y]$ is equal to the answer to $\textsc{Count}(x,y+n)$. Thus, we can build a 2D range counting data structure in $O(n+T(2n,n))$ time that answers queries in $O(1+Q(2n,n))$ time.
\end{proof}

\begin{proof}[Proof of \Cref{lemma:A1A2}]
Let $\texttt{parent}'(u)$ be the parent node of a given node $u$ in $\texttt{Trie}(T[l, r])$, we have
\begin{align*}
    \textsc{CountDistinct}(l, r) &= \sum_{u \text{ is a non-root node of } \texttt{Trie}(T[l, r])} \mathds{1}\{\texttt{str}'(u) \in \mathcal{D}\} \\
    &= \sum_{u \text{ is a non-root node of } \texttt{Trie}(T[l, r])} (f'_u - f'_{\texttt{parent}'(u)}),
\end{align*}
which is equal to $A_1 - A_2$ by considering the coefficient of each $f'_u$.
\end{proof}

\begin{proof}[Proof of \Cref{lemma:A1}]
    There exists a bijection relationship between $\{T[i, r] \mid l \leq i \leq \pos\}$ the set of all leaf nodes in $\texttt{Trie}(T[l, r])$. This is because: 
    
    \textbf{(1)} For each string in $\{T[i, r] \mid l \leq i \leq \pos\}$, the node in $\texttt{Trie}(T[l, r])$ that represents it is definitely a leaf. If not, suppose the substring that corresponds to a non-leaf node is $T[i, r]$, which implies that $T[i, r]$ is a strict prefix of another $T[l, r]$'s substring, denoted as $T[l_1, r_1], (l \leq l_1 \leq r_1 \leq r)$. However, this indicates that $T[\pos, r]$ is a strict prefix of $T[l_1 + \pos - i, r_1]$, meaning that $T[\pos,r]$ has another occurrence in $T[l,r]$, contradicting to \Cref{def:pos}; 
    
    \textbf{(2)} For each leaf node $u$ in $\texttt{Trie}(T[l, r])$, it can not represent a substring of $T[i, j]$ with $j < r$ ($T[i, j + 1]$'s corresponding node can be one of $u$'s child), or $i > \pos$ (\Cref{def:pos}).

    Therefore, we have:
    \begin{align*}
        A_1 &= \sum\limits_{u \text{ is a leaf node of } \texttt{Trie}(T[l, r])} f'_u \\ 
        &= \sum\limits_{u \text{ is a leaf node of } \texttt{Trie}(T[l, r])} \  \sum_{s \text{ is a prefix of } \texttt{str}'(u)} \mathds{1}\{s \in \mathcal{D}\} \\
        &= \sum_{i = l}^{\pos} \sum_{j = i}^{r} \mathds{1}\{T[i, j] \in \mathcal{D}\} = \sum_{i = l}^{r} \sum_{j = i}^{r} \mathds{1}\{T[i, j] \in \mathcal{D}\} - \sum_{i = \pos + 1}^{r} \sum_{j = i}^{r} \mathds{1}\{T[i, j] \in \mathcal{D}\} \\
        &= \textsc{Count}(l,r)-\textsc{Count}(\pos + 1, r)
    \end{align*}
\end{proof}

\begin{proof}[Proof of \Cref{lemma:A2}]
Consider mapping each node \( u \) in \(\texttt{Trie}(T[l, r])\) with \( \texttt{cnt}'_u \geq 2 \) to a node \( p(u) \) in \(\texttt{PreTree}(T) \) such that \(\texttt{str}'(u) \in \texttt{str}(p(u)) \). Given \( \cnt'_u \geq 2 \), \(\texttt{str}'(u)\) must be the longest string \(\texttt{maxstr}(p(u))\) in \(\texttt{str}(p(u))\). By recalling the definitions of \( \cnt \) and \( f \), this further implies that \( \cnt_{p(u)} = \cnt'_u \) and \( f_{p(u)} = f'_u \).

Additionally, for each node \( v \) in \(\texttt{PreTree}(T) \) with \( \cnt_v \geq 2 \), we can map \( v \) to a node \( q(v) \) in \(\texttt{Trie}(T[l, r]) \) such that \(\texttt{str}'(q(v)) = \texttt{maxstr}(v) \).
Thus, there exists a bijection between \(S_1 = \{ u \mid u \in \texttt{Trie}(T[l, r]), \cnt'_u \geq 2 \}\) and \(S_2 = \{ v \mid v \in \texttt{PreTree}(T), \cnt_v \geq 2 \}\), where for each node $u \in S_1$ and its corresponding node $v \in S_2$, we have $f'_u = f_v$ and $\cnt'_v = \cnt_v$.

Given the above bijection relationship, we have
\begin{align*}
    A_2 &= \sum\limits_{u \text{ is a non-leaf node of } \texttt{Trie}(T[l, r])} (\cnt'_u - 1) f'_u \\
    &= \sum\limits_{u \in \texttt{Trie}(T[l, r]), \cnt'_u \geq 1} (\cnt'_u - 1) f'_u \\
    &= \sum_{u\in \texttt{PreTree}(T), \cnt_u \geq 1} (\cnt_u-1)f_u
\end{align*}

\end{proof}

\begingroup
\emergencystretch=1em
\sloppy
\begin{proof}[Proof of \Cref{lemma:p1-string-to-tree}] 
    As discussed above, $\pos$ is the largest number that has $b^{(r-1)}(T[\pos,r])<l+r-\pos$. Assume $T[i,r]\in \texttt{str}(y)$, then the constriant becomes $\pos\leq r+l-a^{(r-1)}(y)-1$. Additionally, $T[\pos,r]\in \texttt{str}(y)$ if and only if $\texttt{len}(\texttt{parent}(y))<r-\pos+1\leq \texttt{len}(y)$. Thus, such a value $\pos$ exists only when $r-\texttt{len}(y)+1\leq r+l-a^{(r-1)}(y)-1$, which is $a^{(r-1)}(y)-\texttt{len}(y)+1<l$. Since we require $T[\pos,r]$ to be as short as possible, we want $y$ to be as low as possible, and $\pos$ should be as large as possible. Thus, we have
    \[
    \begin{aligned}
    \pos=\min\{&r+l-a^{(r-1)}(y)-1,\\
    &r-\texttt{len}(\texttt{parent}(y))\}.
    \end{aligned}
    \]
\end{proof}
\endgroup

\begingroup
\emergencystretch=1em
\begin{proof}[Proof of \Cref{lemma:event-seq}]
    Assume we have obtained sets $E_1,E_2,\dots,E_{r-1}$, as well as $a^{(r-1)}(u)$ for all $u\in \texttt{SufTree}(T)$ by a link-cut tree, where we assume all vertices $x$ in each solid path have the same value of $a^{(r-1)}(x)$. Now consider finding the set $E_r$.
    
    We need to assign $a^{(r)}(x) = r$ for all vertices $x$ in path from $x_r$ to the root, and keep the value of $e^{(r-1)}(x)$ for other vertices.
    To do so, we create a new solid path by $\texttt{expose}(x_r)$ from the link-cut tree technique and assign $e^{(r)}(x) = r$ for all vertices in the path.
    After that, all vertices $x$ in each solid path will still have the same value of $e^{(r)}(x)$.
    
    Considering all solid paths before $\texttt{expose}(x_r)$ that has an intersection with the path from $x_r$ to the root, we denote the endpoints of those intersections as $I_r = \{(u_1, v_1), \ldots, (u_{|I_r|}, v_{|I_r|})\}$.
    For each $(u, v) \in I_r$, without loss of generality, let $u$ have the smaller depth.
    We have $v$ either equals $u$ or is $u$'s ancestor, and all vertices $x$ in the path from $u$ to $v$ have the same value of $a^{(r-1)}(x)$ since they are in the same solid path. 
    Therefore, all $(r, a^{(r-1)}(u), u, v)$s are valid events, and we add them to $E_r$ in the order of $u$'s depth, which is exactly the order of visiting them during $\texttt{expose}(x_r)$.

    Totally, we have $\sum_{i=1}^n |I_i|$ number of events. Given that the number of solid paths that have an intersection with a new solid path of $\texttt{expose}(x_i)$ is amortized $O(\log n)$, the size of the constructed event sequence is $O(n \log n)$. Also, since the cost of $\texttt{expose}$ operation is $O(\log n)$ amortized time, these sets of events can be constructed in $O(n\log n)$ time.
    
\end{proof}
\endgroup

\begin{proof}[Proof of \Cref{lemma:increment-condition}]
    The shortest string in $\texttt{str}(x)$ appears in $T[l,n]$ firstly at $T[l, l+\texttt{len}(\texttt{parent}(x))]$, and appears in $T[l+1,n]$ firstly as $T[j,j+\texttt{len}(\texttt{parent}(x))]$. So $x$ causes $\cnt_{\texttt{parent}(x)}$ to increases if and only if the shortest string in $\texttt{str}(x)$ appears in $T[l,r]$ but not in $T[l+1,r]$, which is equivalent to $r\in [i + \texttt{len}(\texttt{parent}(x)), j + \texttt{len}(\texttt{parent}(x)))$.
\end{proof}

\begin{proof}[Proof of \Cref{lemma:one-in-A2}]
For all $r\in [i + \texttt{len}(y), j + \texttt{len}(y))$, strings in $\texttt{str}(y)$ occur at most twice in $T[l, r]$, i.e., one starts from $i$ and possibly another starts from $j$.
The first occurrence corresponds to the child $x$, and the second occurrence cannot be extended to a child of $y$ because there is no enough space for the string to be extended to length $\geq \texttt{len}(y)+1$. Therefore, $\cnt_{y} = 1$.
\end{proof}

\begin{proof}[Proof of \Cref{lemma:cnt_more_two_hold}]
    The condition $r<j_0+\texttt{len}(u_1)$ is because of the increment condition \cref{lemma:increment-condition}. Now we check the lower bound of $r$. When $r<j_1+\texttt{len}(u_1)$, by following the same argument in the proof of \cref{lemma:one-in-A2}, we know that $\cnt_{u_1}=1$. If $j_0=j_1$, the interval becomes empty and we have already finished the proof. Now we assume $j_0\neq j_1$. When $r\geq j_1+\texttt{len}(u_1)$, the string $\texttt{maxstr}(u_1)$ appears at least twice in $T[l,r]$, i.e., one starts from $i$ and another starts from $j_1$. The first occurrence corresponds to the child $v_0$, and the second occurrence will lead to another child of $u_1$ since there is enough space ($r\geq j_1+\texttt{len}(u_1)$) and $j_0\neq j_1$. Thus, we have $\cnt_{u_1}\geq 2$.
\end{proof}